\documentclass{article}

\usepackage{arxiv}

\usepackage[utf8]{inputenc} 
\usepackage[T1]{fontenc}    
\usepackage{hyperref}       
\usepackage{url}            
\usepackage{booktabs}       
\usepackage{amsfonts}       
\usepackage{nicefrac}       
\usepackage{microtype}      
\usepackage{lipsum}
\usepackage{natbib}
	\usepackage{upgreek}
	\usepackage{multirow}
	\usepackage{verbatim}
	\usepackage{rotating}
	\usepackage{graphicx}
	\usepackage{listings}
	\usepackage{threeparttable}
	\usepackage{float}
	\usepackage{color}
	\usepackage{amsmath}
	\usepackage[caption = false]{subfig}
	\usepackage{chngcntr}

\title{A Joint Bayesian Space-Time Model to Integrate Spatially Misaligned Air Pollution Data in R-INLA}

\author{
  C.~Forlani\thanks{Correspondence to: \texttt{c.forlani@imperial.ac.uk}} \\
  MRC Centre for Environment and Health \\ 
  Department of Epidemiology and Biostatistics \\
  School of Public Health \\
  Imperial College London \\
  London, UK \\
   \And
 S.~Bhatt \\
  Department of Infectious Disease Epidemiology \\
  School of Public Health \\
  Imperial College London \\
  London, UK \\
     \And
      M.~Cameletti \\
     Department of Management \\
     Economics and Quantitative Methods \\
     Universit\`a degli Studi di Bergamo \\
     Bergamo, Italy\\
     \And
 E.~Krainski \\
  Department of Statistics \\
  Universidade Federal do Paran\'a \\
  Paran\'a, Brazil
     \And
 M.~Blangiardo \\
  MRC Centre for Environment and Health \\ 
  Department of Epidemiology and Biostatistics \\
  School of Public Health \\
  Imperial College London \\
  London, UK \\
}

\begin{document}
\maketitle

\begin{abstract}
In air pollution studies, dispersion models provide estimates of concentration at grid level covering the entire spatial domain, and are then calibrated against measurements from monitoring stations. However, these different data sources are misaligned in space and time. If misalignment is not considered, it can bias the predictions. 
We aim at demonstrating how the combination of multiple data sources, such as dispersion model outputs, ground observations and covariates, leads to more accurate predictions of air pollution at grid level.
We consider nitrogen dioxide (NO$_2$) concentration in Greater London and surroundings for the years 2007-2011, and combine two different dispersion models. Different sets of spatial and temporal effects are included in order to obtain the best predictive capability.
Our proposed model is framed in between calibration and Bayesian melding techniques for data fusion red. Unlike other examples, we jointly model the response (concentration level at monitoring stations) and the dispersion model outputs on different scales, accounting for the different sources of uncertainty. 
Our spatio-temporal model allows us to reconstruct the latent fields of each model component, and to predict daily pollution concentrations.
We compare the predictive capability of our proposed model with other established methods to account for misalignment (e.g. bilinear interpolation), showing that in our case study the joint model is a better alternative. 
\end{abstract}


	\section{Introduction} \label{s:background}
	
	Air pollution is a major concern for policy makers worldwide \citep{ECAQS,EPANAAQS,WHOguidelines}, and there is extensive evidence of its negative effects, in particular on respiratory and cardiovascular diseases \citep{Dominici2010,Comeap2015,Atkinson2015, Lipfert2017}. Obtaining an accurate estimate of air pollution concentration is key for evaluating compliance with regulatory standards set by national and international environmental agencies and to reduce exposure misclassification in epidemiological studies \citep{Berrocal2012,shaddick2017,Keller2019}.
	
	Air pollution data come from different sources, each presenting some limitations: ground measurements from monitoring network stations, usually affected by sparse spatial resolution; estimates from Land Use Regression models (LUR), which rely on the availability of accurate and dense monitor observations; satellite remote sensing data, sometimes poorly correlated with ground pollution level; simulations from deterministic models (e.g. chemical transport models or dispersion models), that can present prediction quality concerns despite the complete spatial coverage and high temporal resolution \citep{Shaddick2002,Lee2011a,Gelfand2012,Shaddick2015,Hoek2008,Johnson2010,chang2016,shaddick2017}.
	
	Several `hybrid' appproaches have been proposed to combine these data sources to draw from their strengths and to overcome their limitations, but not all of them address the discrepancy in the spatial resolution of the different data sources, which is known as {\it misalignment} or {\it change of support problem} (COSP).

	\subsection{Main approaches to address spatial misalignment} \label{s:lit_rev}
	In the context of COSP, we refer to upscaling methods when the target resolution is lower than the data resolution (e.g. point-to-area), and to downscaling when the target resolution is higher (e.g. area- or grid-to-point). 
	
	Model-based solutions for {\it data assimilation} (also referred to as {\it data fusion} or {\it data blending}) in environmental applications allow us to account for all sources of uncertainty while addressing COSP. These are usually set within a hierarchical Bayesian framework. Popular approaches include Bayesian melding and calibration techniques \citep{chang2016}. 
	
	Bayesian melding assumes that both measurements and modelled data are error-prone realizations of an underlying latent true pollution field, and they both inform the posterior distribution of the latent process. Among the proposed melding strategies applied to misaligned air pollution data we find, for instance, the downscaling spatial Bayesian melding model by \citet{Fuentes2005}, the upscaling spatial Bayesian melding model by \citet{Wikle2005}, and the upscaling spatio-temporal fusion model by \citet{McMillan2010}.
	
	Calibration techniques assume that the model-based estimates (e.g. from dispersion models) are used in a regression framework as predictors against the monitoring site measurements. In this way the computational cost is reduced compared to melding, as the models only need to be fitted at the monitoring sites locations \citep{Berrocal2012,chang2016}. Some examples are the block-averaging upscaling calibration fusion model by \citet{Sahu2010}, and the spatio-temporal downscaling calibration models by \citet{Berrocal2010,Berrocal2012}, which can be considered a generalization of a Bayesian universal kriging model \citep{Gelfand2019}.

	\subsection{Novelty of our approach} \label{s:intro}
	
	In this paper, we are framed in the context of data integration to improve air pollution predictions at a fine grid. We combine monitoring measurements and numerical model outputs coming from two dispersion models, the Pollution Climate Mapping (PCM) from DEFRA \citep{DEFRA,ricardo2017} model and the Air Quality Unified Model (AQUM) from the Met Office \citep{Savage2013,AQUM}, and account for their associated errors. 
	
	These deterministic models have previously been used for similar purposes: \citet{Lee2015} provide an example of point-to-area upscaling from the PCM model grid to local authority areas for epidemiological applications and \citet{Mukhopadhyay2017} combine the AQUM and the monitoring observations to accurately predict NO$_2$ concentration in UK.
		
	However, usually in the literature only one extra data source at a time is considered \citep{Wikle2005,Fuentes2005,McMillan2010,Sahu2010,Berrocal2010,Berrocal2012,Zidek2012,Huang2015,Lee2015,Pannullo2016,Mukhopadhyay2017,Lee2017,Huang2017,Moraga2017}. We show that, when more are available, these can all be put together to get better predictions while accounting for the bias which affects deterministic data.
	
	Our approach is similar to the coregionalization model proposed by \citet{Schmidt2003} to model CO, NO, and NO$_2$ which allows us to calibrate the deterministic models against the monitor observations through a coefficient similarly to calibration techniques. However, here we treat the three sources of information on NO$_2$ as coming from the same true underlying spatio-temporal process (i.e. the true air pollution concentration field) as in Bayesian melding. The pure application of this kind of models is computationally prohibitive for the high resolution output data we have at hand. This issue is solved by representing the spatially continuous fields as solutions to a Stochastic Partial Differential Equation (SPDE) to handle this in a computationally efficient way \citep{Lindgren2011,krainski2019}.
	
	Additionally, our model reconstructs the continuous latent spatial and temporal fields allowing us to account for all the sources of uncertainty: first, the one associated with the estimates from the numerical models, which is not provided as they are deterministic models, and second, the measurement error associated with ground observations. This is most useful in the perspective of using the predictions from the air pollution model as a measure of exposure in an epidemiological model, where the uncertainty could be fed forward (see for example \citealt{Lee2017} and \citealt{Cameletti2019}). 
	
	The inference is done under the Bayesian paradigm through the Integrated Nested Laplace Approximations (INLA) coupled with the SPDE approach, which is implemented in the \texttt{R-INLA} package \citep{inlaproject}. 
	
	Other authors have implemented solutions for spatially misaligned air pollution data in \texttt{R-INLA}, however their approaches differ from ours under several points of view. In particular, \citet{Moraga2017} show an example of area-to-point misalignment addressed via block averaging, in a spatial-only context, without accounting for the uncertainty associated with the raster data. \citet{Cameletti2019} implement a spatial upscaler from point to area comparing two different averaging methods. \citet{Kifle2017} compare additive and coupled spatio-temporal processes for multivariate data in a biological context (prevalence of vectors for arboviruses), where the data are not misaligned, and do not include any explanatory covariate.
	
	To the best of our knowledge, this is the first time a spatio-temporal model for spatially misaligned point-referenced data is implemented through the INLA-SPDE approach considering more than one deterministic model output at different spatial and temporal resolutions. 

	We compare and contrast several models and through a cross-validation method we evaluate which one produces the most accurate predictions of NO$_2$ concentration in Greater London and surroundings for the period 2007 to 2011. We compare our method with two approaches in which the alignment is done through bilinear interpolation or kriging, hence not accounting for the measurement error associated with the misaligned covariates. The first is a simple hierarchical model that includes linear effects for the covariates and structured spatio-temporal residuals. The second is the recently proposed data integration model from \citet{Mukhopadhyay2017}, which allows for non-stationarity in the residual spatial process.
	
	The remainder of the article is organised as follows: section \ref{s:data} presents the study area and data; section \ref{s:methods} describes the methods used in the analysis, starting with the model specification followed by a description of the competitor models; section \ref{s:results} reports the results of the application of such methods to our air pollution data; finally, section \ref{s:conclusion} contains the conclusions and a short discussion, pointing towards further developments.

	\section{Study area and data} \label{s:data}
	The study focuses on NO$_2$, as it is one of the pollutants regulated by national and international directives, and it is traffic driven, hence characterised by high spatio-temporal variability.
		
	We used daily averages of hourly observations from different monitoring networks including the AEA and the Authomatic Urban and Rural Network (AURN) from the DEFRA's UK Air Quality Archive, and the London Air Quality Network (LAQN) in Greater London and surroundings, managed by the King's College London Environmental Research Group (ERG). The combined database was built as part of the Spatio/Temporal Exposure Assessment Methods (STEAM) project \citep{STEAMweb}. 
	
	We also considered the outputs of two deterministic models: (i) annual 1km$\times$1km predictions from the PCM model provided by DEFRA \citep{DEFRA}, for 2007-2011; (ii) daily 12km$\times$12km predictions from the AQUM model, available for the years 2007-2011, provided by the Met Office \citep{AQUM}. 
	
	We consider the period 2007-2011 due to the availability of the AQUM data.
	
	Among the 213 monitoring stations active between 01/01/2007 and 31/12/2011 for at least 1370 consecutive days (75\% of the total number of days), 126 have been included in the analysis after applying the following criteria to the NO$_2$ time series: (i) the daily average is computed only for the days where at least 18 hourly observations, i.e. the 75\%, are present; (ii) we eliminated non-positive daily averages which do not allow for the logarithmic transformation required in the analysis (negative observations are due to measurement error); (iii) monitors where the resulting daily NO$_2$ is available for less than 1370 days, not necessarily consecutive, have been excluded (this is because an active monitor does not necessarily record NO$_2$ measurements).
	
	The monitors are split into 6 groups maximizing similarity criteria between groups, and a 6-fold cross validation is performed.
	
	For each monitor we have information about the site type classification, that we aggregated into 3 categories: rural, urban, road-kerb side.
	
	We define our study area as that including all the selected monitors, containing 495 grid cells for AQUM and 44,117 grid cells for PCM.

\begin{figure}
		\centerline{\includegraphics[width=.8\textwidth]{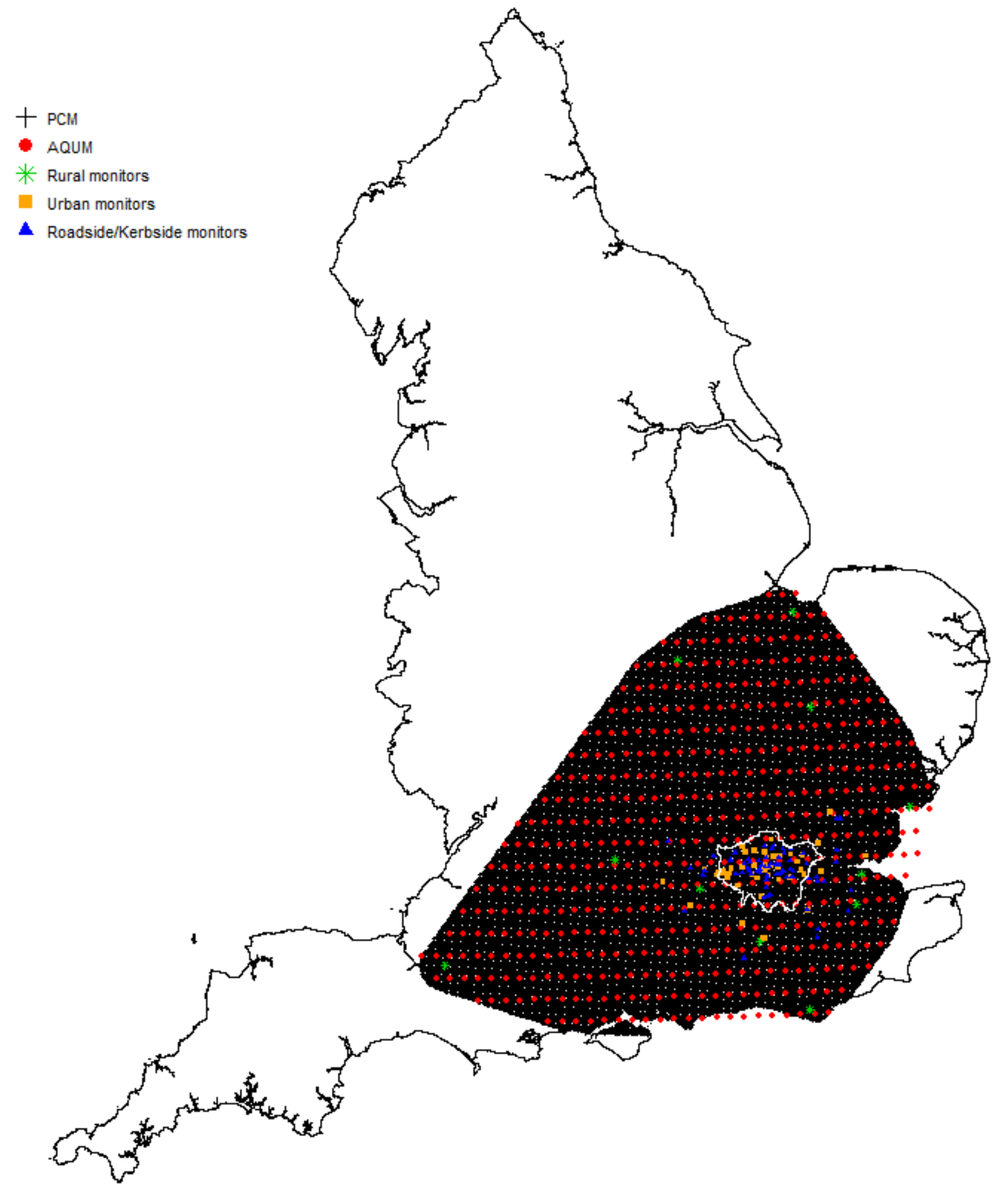}}
		\caption{NO$_2$ data locations on the study domain, on the shape of England as reference.}
		\label{f:data}
	\end{figure}

	The locations of the air pollution data sources described above are displayed in Figure~\ref{f:data}.
	
	The AQUM model includes chemistry, physical and aerosol models, meteorological configuration based on the Met Office’s North Atlantic and European Model (NAE) and emission data \citep{Savage2013}; the PCM model input includes emission inventory, energy projections, road traffic counts, road transport activity and meteorological hourly data from Waddington weather station \citep{ricardo2017}.
	
	Although data were available, we decided against the inclusion of meteorological variables in our models, as they are already an input for both the numerical models considered in the analysis. 
	
	Consistently with the selection criteria, all the 6 training and validation sets have similar distribution of daily NO$_2$ concentration by site type (see Appendix~\ref{a:figures}, Fig.~\ref{f:descriptive}). As expected, in all sets the road-kerb side monitors have higher mean and maximum levels of NO$_2$.
	
	In particular, 17 road-kerb side sites overcome the limits set by the WHO and the European Commission for the annual average of 40$\mu g/m^3$, for at least 4 of the 5 years under study (see Appendix~\ref{a:figures}, Fig.~\ref{f:exceedance}). Of these, the monitor in Lambeth-Brixton Road (LB4) is also well above the threshold of 18$\mu g/m^3$ not to be exceeded more than 18 times annually, for every year, even though a decreasing trend can be observed (from 865 hourly exceedances in 2007, to 62 in 2011), and other 6 monitors exceeded this threshold between 2007 and 2008.
		
		Table~\ref{t:descriptive_data} 
		reports summary statistics for PCM data by year, AQUM data by month and monitor observations by site type. 
		
		\begin{table} 
			\centering
			\begin{threeparttable}
				\small{
					\caption{Descriptive statistics of NO$_2$ concentration ($\mu g/m^3$) for the three data sources \label{t:descriptive_data}}
					\label{t:mod_comp}
					\begin{tabular}{p{0.2cm}p{0.1cm}p{0.6cm}p{0.6cm}p{0.6cm}p{0.6cm}p{0.6cm}p{0.6cm}p{0.6cm}p{0.6cm}}
						\noalign{\smallskip}\hline\noalign{\smallskip}                                                                                            Data    						 		
						&& \multicolumn{1}{r}{Min} 
						&  \multicolumn{1}{r}{1st Qu.}
						&  \multicolumn{1}{r}{Median} 
						&  \multicolumn{1}{r}{Mean}
						&  \multicolumn{1}{r}{3rd Qu.}
						& \multicolumn{1}{r}{Max}  \\
						\noalign{\smallskip}\hline\noalign{\smallskip} 				
						\multicolumn{1}{l}{PCM 2007}   &&   \multicolumn{1}{r}{7.276} & \multicolumn{1}{r}{ 9.948} & \multicolumn{1}{r}{11.289} & \multicolumn{1}{r}{13.039} & \multicolumn{1}{r}{14.055} & \multicolumn{1}{r}{66.036} \\ 
						\multicolumn{1}{l}{PCM 2008}   &&   \multicolumn{1}{r}{5.615} & \multicolumn{1}{r}{ 9.107} & \multicolumn{1}{r}{10.439} & \multicolumn{1}{r}{12.064} & \multicolumn{1}{r}{13.198} & \multicolumn{1}{r}{56.123} \\ 
						\multicolumn{1}{l}{PCM 2009}   &&   \multicolumn{1}{r}{7.097} & \multicolumn{1}{r}{10.105} & \multicolumn{1}{r}{11.333} & \multicolumn{1}{r}{12.844} & \multicolumn{1}{r}{13.915} & \multicolumn{1}{r}{53.290} \\ 
						\multicolumn{1}{l}{PCM 2010}   &&   \multicolumn{1}{r}{5.613} & \multicolumn{1}{r}{10.396} & \multicolumn{1}{r}{11.843} & \multicolumn{1}{r}{13.404} & \multicolumn{1}{r}{14.705} & \multicolumn{1}{r}{62.190} \\ 
						\multicolumn{1}{l}{PCM 2011}   &&   \multicolumn{1}{r}{5.437} & \multicolumn{1}{r}{ 9.905} & \multicolumn{1}{r}{11.317} & \multicolumn{1}{r}{12.545} & \multicolumn{1}{r}{13.575} & \multicolumn{1}{r}{55.400} \\ 
						\hline 
						\multicolumn{1}{l}{AQUM January}   && \multicolumn{1}{r}{0.215} & \multicolumn{1}{r}{11.354} & \multicolumn{1}{r}{19.926} & \multicolumn{1}{r}{21.943} & \multicolumn{1}{r}{30.150} & \multicolumn{1}{r}{ 73.654} \\ 
						\multicolumn{1}{l}{AQUM February}  && \multicolumn{1}{r}{0.154} & \multicolumn{1}{r}{13.912} & \multicolumn{1}{r}{22.491} & \multicolumn{1}{r}{25.052} & \multicolumn{1}{r}{33.535} & \multicolumn{1}{r}{ 97.435} \\ 
						\multicolumn{1}{l}{AQUM March}     && \multicolumn{1}{r}{0.291} & \multicolumn{1}{r}{10.165} & \multicolumn{1}{r}{16.760} & \multicolumn{1}{r}{19.180} & \multicolumn{1}{r}{25.969} & \multicolumn{1}{r}{ 84.589} \\ 
						\multicolumn{1}{l}{AQUM April}     && \multicolumn{1}{r}{0.059} & \multicolumn{1}{r}{ 9.982} & \multicolumn{1}{r}{15.339} & \multicolumn{1}{r}{17.248} & \multicolumn{1}{r}{22.457} & \multicolumn{1}{r}{ 81.498} \\ 
						\multicolumn{1}{l}{AQUM May}   	   && \multicolumn{1}{r}{0.008} & \multicolumn{1}{r}{ 6.847} & \multicolumn{1}{r}{10.340} & \multicolumn{1}{r}{12.187} & \multicolumn{1}{r}{15.622} & \multicolumn{1}{r}{ 77.623} \\ 
						\multicolumn{1}{l}{AQUM June}      && \multicolumn{1}{r}{0.000} & \multicolumn{1}{r}{ 6.493} & \multicolumn{1}{r}{ 9.568} & \multicolumn{1}{r}{11.088} & \multicolumn{1}{r}{13.974} & \multicolumn{1}{r}{ 63.689} \\ 
						\multicolumn{1}{l}{AQUM July}      && \multicolumn{1}{r}{0.409} & \multicolumn{1}{r}{ 5.855} & \multicolumn{1}{r}{ 8.388} & \multicolumn{1}{r}{ 9.842} & \multicolumn{1}{r}{12.244} & \multicolumn{1}{r}{ 67.893} \\ 
						\multicolumn{1}{l}{AQUM August}    && \multicolumn{1}{r}{0.000} & \multicolumn{1}{r}{ 6.392} & \multicolumn{1}{r}{ 9.210} & \multicolumn{1}{r}{10.591} & \multicolumn{1}{r}{13.176} & \multicolumn{1}{r}{ 58.456} \\ 
						\multicolumn{1}{l}{AQUM September} && \multicolumn{1}{r}{0.054} & \multicolumn{1}{r}{ 7.545} & \multicolumn{1}{r}{11.795} & \multicolumn{1}{r}{13.899} & \multicolumn{1}{r}{18.095} & \multicolumn{1}{r}{ 68.396} \\ 
						\multicolumn{1}{l}{AQUM October}   && \multicolumn{1}{r}{0.067} & \multicolumn{1}{r}{10.521} & \multicolumn{1}{r}{16.622} & \multicolumn{1}{r}{18.402} & \multicolumn{1}{r}{24.576} & \multicolumn{1}{r}{ 70.455} \\ 
						\multicolumn{1}{l}{AQUM November}  && \multicolumn{1}{r}{0.531} & \multicolumn{1}{r}{11.714} & \multicolumn{1}{r}{18.694} & \multicolumn{1}{r}{20.942} & \multicolumn{1}{r}{28.109} & \multicolumn{1}{r}{ 97.137} \\ 
						\multicolumn{1}{l}{AQUM December}  && \multicolumn{1}{r}{0.000} & \multicolumn{1}{r}{12.890} & \multicolumn{1}{r}{23.150} & \multicolumn{1}{r}{25.040} & \multicolumn{1}{r}{34.480} & \multicolumn{1}{r}{121.320} \\ 
						\hline
						\multicolumn{1}{l}{Monitors RUR}  	&&   \multicolumn{1}{r}{0.484} & \multicolumn{1}{r}{ 5.357} & \multicolumn{1}{r}{ 8.995} & \multicolumn{1}{r}{11.405} & \multicolumn{1}{r}{14.734} & \multicolumn{1}{r}{114.375} \\ 
						\multicolumn{1}{l}{Monitors URB}  	&&   \multicolumn{1}{r}{0.679} & \multicolumn{1}{r}{11.938} & \multicolumn{1}{r}{19.091} & \multicolumn{1}{r}{22.286} & \multicolumn{1}{r}{29.049} & \multicolumn{1}{r}{166.792} \\ 
						\multicolumn{1}{l}{Monitors RKS}  	&&   \multicolumn{1}{r}{0.245} & \multicolumn{1}{r}{20.208} & \multicolumn{1}{r}{30.524} & \multicolumn{1}{r}{34.903} & \multicolumn{1}{r}{44.422} & \multicolumn{1}{r}{231.292} \\ 
						\noalign{\smallskip}\hline\noalign{\smallskip}
				\end{tabular}}
			\end{threeparttable}
		\end{table}

	\section{Methods} \label{s:methods}	
	
	In this section we first present some analysis on the AQUM and PCM data, then we introduce the joint model, and finally the models that we use for comparison. Note that we will represent vector/matrices in bold typeface.
	
		\subsection{Separate models for AQUM and PCM data} \label{s:separate_models} 
	
	In order to quantify the relevance of the temporal component for PCM ($i=1$) and the spatial component for AQUM ($i=2$), we ran three models for each data source separately: (i) one with spatial-only or temporal-only effect respectively, (ii) one with additive spatial and temporal effects and (iii) one with a spatio-temporal interaction. 
	
	Let's define $\boldsymbol{y_i}$ as the vector of air pollution concentration on the logarithmic scale across space and time for the $i$-th numerical model. This is assumed to be normally distributed with mean $\boldsymbol{\eta_i}$ and variance $\sigma^{2}_{\epsilon_i}$: $\boldsymbol{y_i} \sim MVN(\boldsymbol{\eta_i}, \sigma^{2}_{\epsilon_i}I) $
	
	Each element of the linear predictor $\boldsymbol{\eta_i}$ (for a time point $t$ and location $\boldsymbol{s}$ identified by UTM coordinates) for models (i), (ii), (iii) is specified as follows:
	\begin{equation} \tag{i} \label{eq:linpred_separate_models}
	\begin{split} 
	\eta_1(\boldsymbol{s}) = \alpha_1 + z_{11}(\boldsymbol{s})  \\
	\eta_2(t) = \alpha_2 + z_{22}(t)   
	\end{split}
	\end{equation}
	\begin{equation} \tag{ii} 
	\begin{split} 
	\eta_1(\boldsymbol{s},t) = \alpha_1 + z_{11}(\boldsymbol{s}) + z_{21}(t)  \\
	\eta_2(\boldsymbol{s},t) = \alpha_2 + z_{12}(\boldsymbol{s}) + z_{22}(t)    
	\end{split}
	\end{equation}  
	\begin{equation} \tag{iii} 
	\begin{split} 
	\eta_1(\boldsymbol{s},t) = \alpha_1 + z_{31}(\boldsymbol{s},t)    \\
	\eta_2(\boldsymbol{s},t) = \alpha_2 + z_{32}(\boldsymbol{s},t)     
	\end{split}
	\end{equation}
	where $z_{1i}(\boldsymbol{s})$ is the realisation at location $\boldsymbol{s}$ of the spatial process $\boldsymbol{z_{1i}}$ with Mat\'ern covariance function $\boldsymbol{z_{1i}} \sim MVN(\textbf{0}, \sigma^2_{z_{1i}}\boldsymbol{\Sigma}) $, $z_{2i}(t) \sim N(z_{2i}(t-1), \sigma^2_{z_{2i}}) $ is a temporal process modelled as a random walk and $\boldsymbol{z_{3i}} \sim MVN(\textbf{0}, \sigma^2_{z_{3i}}\boldsymbol{\Sigma_t} \otimes \boldsymbol{\Sigma_s}) $ is a separable space-time interaction with Mat\'ern covariance function and temporal dependence modelled as a random walk.
	
	Based on the deviance information criterion (DIC), the results show that AQUM spatial variation is relevant but there is no need for a space-time interaction so model (ii) is selected for AQUM, while the PCM temporal variation is negligible so model (i) is selected for PCM (see Appendix~\ref{a:separate_models} for details). Hence, we will use this specification in the joint model presented in the next section.

	\subsection{Bayesian joint spatio-temporal model for misaligned covariates} \label{s:model}
	
	Following \citet{Kifle2017} we implement an additive space-time model for data observed at different points in space, which share a spatial and a temporal component. 
		
	Previous similar applications consider measurements of more than one variable at the same locations, but this is not a requirement in the INLA-SPDE approach. 
	
	Our model is joint in the sense that we specify one likelihood for the response and one for each of the misaligned covariates, and they contain common components which are estimated using all the data. Even though in \texttt{R-INLA} the problem is computationally treated similarly to a multivariate situation, this is not our case as we ultimately consider solely the monitor observations as response variable.
	
	We make the assumption that the same temporal dynamics govern AQUM and monitor observations, and likewise the same spatial dynamics govern PCM, AQUM and monitor observations. 
	
	Our hierarchical model has three levels: in the first we define the likelihoods, in the second the random effect components, while the third level includes the prior distributions for the model parameters and hyperparameters.
	
	The joint model presented below is implemented via INLA, a computationally efficient alternative to Markov chain Monte Carlo (MCMC) methods that works specifically on hierarchical Gaussian Markov Random Fields (GMRF). Details on how this is done in \texttt{R-INLA} can be found in Appendix~\ref{a:joint_inla}.			
	
	\subsubsection{Level 1: Likelihoods and linear predictors}
	
	Let $y_i(\boldsymbol{s},t)$ denote the PCM ($i=1$) and AQUM ($i=2$) data and the observed NO$_2$ concentration ($i=3$) at the generic time point $t$ and site $\boldsymbol{s}$, on the logarithmic scale. These are assumed to be normally distributed, with mean $\eta_i(\boldsymbol{s},t)$ and measurement error variance $\sigma^2_{\epsilon_i}$: 
	
	$y_1(\boldsymbol{s},t) \sim N(\eta_1(\boldsymbol{s}), \sigma^{2}_{\epsilon_1}) \qquad $(PCM)
	
	$y_2(\boldsymbol{s},t) \sim N(\eta_2(\boldsymbol{s},t), \sigma^{2}_{\epsilon_2}) \qquad $(AQUM)
	
	$y_3(\boldsymbol{s},t) \sim N(\eta_3(\boldsymbol{s},t), \sigma^{2}_{\epsilon_3}) \qquad $(Ground observations)

	Based on the results from section \ref{s:separate_models} we model the PCM data with an intercept and a spatial component and the AQUM data with an intercept and additive spatial and temporal components. These are shared between the three linear predictors, which are the following:
	\begin{equation} \tag{1} \label{eq:linpred1}
	\eta_1(\boldsymbol{s}) = \alpha_1 + z_1(\boldsymbol{s})   
	\end{equation}
	\begin{equation} \tag{2} \label{eq:linpred2}
	\eta_2(\boldsymbol{s},t) = \alpha_2 +  \lambda_{1,2} z_1(\boldsymbol{s}) + z_2(t)   
	\end{equation}  
	\begin{equation} \tag{3} \label{eq:linpred3}
	\eta_3(\boldsymbol{s},t) = \alpha_3 + \beta_{k_s} + \lambda_{1,3} z_1(\boldsymbol{s}) +   \lambda_{2,3} z_2(t) + z_3(t,k{_s})  
	\end{equation}
	where $\alpha_i$ are the intercepts, $\lambda_{i,j}$ are the scaling parameters for the shared components from $\boldsymbol{\eta}_i$ to $\boldsymbol{\eta}_j$, $\beta_{k_s}$ is the fixed effect for the site type as categorical variable ($k_s=0$: rural (reference), $k_s=1$: urban and $k_s=2$: road-kerb side), and $\boldsymbol{z_1}$ and $\boldsymbol{z_2}$ are the shared random effects. The linear predictor for the ground observations $\boldsymbol{\eta}_3$ also contains an interaction term $\boldsymbol{z_3}$ which allows for a different residual temporal trend for each site type.
	
	Note that even though PCM is assumed to be governed only by a spatial effect, its output does vary in both space and time, so the deterministic model output $\boldsymbol{y}_1$ has space and time indices (here the locations are the centroids of the 44117 PCM grid cell, and the time points are the years), while its latent field $\boldsymbol{z_1}$ has only a spatial index. 
	
	For AQUM, the space and time indices of $\boldsymbol{y}_2$ correspond to the centroids of the 495 AQUM grid cell and the 1826 days respectively.
	
	Finally, $\boldsymbol{y}_3$ is measured at the 126 monitors on 1826 days.

	\subsubsection{Level 2: latent fields}
	
	In Equation~(\ref{eq:linpred3}), $\boldsymbol{z_1} \sim MVN(\textbf{0}, \sigma^2_{z_1}\boldsymbol{\Sigma})$ is the common spatial latent field, with $\boldsymbol{\Sigma}$ being the correlation matrix defined by the Mat\'ern stationary and isotropic covariance function (see Appendix \ref{app:matern}). It is important to note that $\boldsymbol{z_1}$ is then rescaled for AQUM and monitor observations through $\lambda_{1,2}$ (eq. \ref{eq:linpred2}) and $\lambda_{1,3}$ (eq. \ref{eq:linpred3}). 
	
	In the same equation, $z_2(t)$ is the $t$-th element of the temporal latent field $\boldsymbol{z_2}$, and is modelled as a random walk: $z_2(t) \sim N(z_2(t-1), \sigma^2_{z_2})$. 
	Similarly to $\boldsymbol{z_1}$, $\boldsymbol{z_2}$ is rescaled for the monitor observations through $\lambda_{2,3}$ (eq. \ref{eq:linpred3}).

	Finally, $\boldsymbol{z_3}$ is the residual temporal trend assumed to be different for each site type (rural, urban, road-kerb side), and modelled as first order autoregressive $z_3(t,k{_s}) \sim N(\rho z_3(t-1,k{_s}), \sigma^2_{z_3})$. In other words, we assume conditionally independent replications of the same latent field for each site type, with shared hyperparameters \citep{Martins2013}.

	\subsubsection{Level 3: priors} \label{s:priors}
	
	The priors on the model parameters are specified as follows.
	
	According to \citet{Fuglstad2017}, we choose a penalised complexity prior \citep{simpson2017} for range and variance of the latent spatial field $\boldsymbol{z_1}$ such that $P(r<r_0)=0.95$ and $P(\sigma_{z_1}>\sigma_0)=0.5$, where $r_0 = 1/5$ of the domain size and $\sigma_0 = 100$ (see Appendix \ref{app:matern}).
	
	For the standard deviation of the random walk we assume a penalised complexity prior such that the probability that $\sigma_{z_2}$ is greater than the empirical standard deviation of the AQUM data is 1\%, i.e. $P(\sigma_{z_2}>SD(AQUM))=0.01$.
	
	For the time-sitetype interaction we assume the default vague prior defined on the log-precision: $log(1/\sigma^2_{z_3}) \sim logGamma(1, 5e-05)$; for the autoregressive parameter we assume $\rho \sim N(0.3,0.5)$ using information from previous modelling exercise.
	
	The precisions of response variable, AQUM data and PCM data are assigned the default vague prior $log(1/\sigma^2_{\epsilon_i}) \sim logGamma(1, 5e-05)$, $i=1,2,3$. 
	
	On the scaling coefficients we put a Normal prior centred on a positive values around 1 with a large variance to ensure minimal information: $\lambda_{1,2} \sim N(1.1, 100)$, $\lambda_{1,3} \sim N(1.3, 100)$,  $\lambda_{2,3} \sim N(0.9, 100)$.
	
	Finally on the coefficients of the fixed effects $\alpha_i$ and $\beta_{k{_s}}$ we assume the default weak Normal prior distribution $N(0,1000)$.
	
	\subsection{Competitor models} \label{s:competitors}
	
	We compared our model to three different competitors: (i) a joint model that includes only one misaligned covariate (either AQUM or PCM), (ii) a simple hierarchical model that includes a covariate aligned at the monitoring sites through bilinear interpolation \citep{Akima1978} or kriging, and (iii) a complex hierarchical model which allows for non-stationarity after interpolating the misaligned covariates via bilinear interpolation \citep{Mukhopadhyay2017}. 
	
	The aim of this comparison is to evaluate if the inclusion of more than one extra data source actually improves the model predictive capability and can counterbalance the need for complex random effect structures, and if there is a gain in moving from a simple interpolation to a modelling framework. 
	
	We describe the three comparators in the rest of this section. 
	
	\subsubsection{Joint models with one misaligned covariate only}	\label{s:onecovar}
	
	The joint model that includes PCM only is specified as follows:
	
	$y_1(\boldsymbol{s},t) \sim N(\eta_1(\boldsymbol{s}), \sigma^{2}_{\epsilon_1}) \qquad $(PCM)
	
	and
	
	$y_2(\boldsymbol{s},t) \sim N(\eta_2(\boldsymbol{s},t), \sigma^{2}_{\epsilon_2}) \qquad $(Ground observations)
	
	with
	
	$\eta_1(\boldsymbol{s}) = \alpha_1 + z_1(\boldsymbol{s})   \qquad \text{(PCM)} $
	
	$\eta_2(\boldsymbol{s},t) = \alpha_2 + \beta_{k_s} + \lambda_{1,2} z_1(\boldsymbol{s}) + z_3(t,k{_s})  \qquad $(Ground observations)
	
	Similarly, the joint model that includes AQUM only is defined as:
	
	$y_1(\boldsymbol{s},t) \sim N(\eta_1(\boldsymbol{s},t), \sigma^{2}_{\epsilon_1}) \qquad $(AQUM)
	
	and
	
	$y_2(\boldsymbol{s},t) \sim N(\eta_2(\boldsymbol{s},t), \sigma^{2}_{\epsilon_2}) \qquad $(Ground observations)
	
	with
	
	$\eta_1(\boldsymbol{s},t) = \alpha_1 +  z_1(\boldsymbol{s}) + z_2(t)   \qquad \text{(AQUM)} $
	
	$\eta_2(\boldsymbol{s},t) = \alpha_2 + \beta_{k_s} + \lambda_{1,2} z_1(\boldsymbol{s}) +   \lambda_{2,2} z_2(t) + z_3(t,k{_s})  \qquad $(Ground observations)
	
	For these models we considered either fixed to 1 or varying calibration coefficients $\lambda_{i,j}$, and different priors. The final choice of priors is the one reported in Section \ref{s:priors}. 
	
	\subsubsection{Data integration model via interpolation} \label{s:bilint}
	
	We implement two models that use interpolation techniques to obtain values of AQUM and PCM at the monitoring stations. The first is a naive bilinear interpolation, the second can be considered as Bayesian kriging, as we predict AQUM and PCM at the monitoring stations from the models described in section \ref{s:separate_models}.
	
	In both cases, after aligning the AQUM ($\boldsymbol{X_1}$) and PCM ($\boldsymbol{X_2}$) values, we consider a linear effect on the covariates, a spatially structured residual $\boldsymbol{z_1}$, a temporally structured residual $\boldsymbol{z_2}$ and the site-type-specific temporal effect $\boldsymbol{z_3}$ specified as in Section \ref{s:model}. We also keep the fixed effects for the site type $\beta_{k_s}$ as in the joint model.
	
	We specify a normal likelihood $y(\boldsymbol{s},t) \sim N(\eta(\boldsymbol{s},t), \sigma^{2}_{\epsilon})$ and the linear predictor as follows:
	
	$\eta(\boldsymbol{s},t) = \beta_0 + \beta_1X_1(\boldsymbol{s},t) + \beta_2X_2(\boldsymbol{s},t) + \beta_{k_s} + z_1(\boldsymbol{s}) + z_2(t) + z_3(t,k{_s})$
	
	\subsubsection{Data integration model with non-stationarity} \label{s:sujit}
	
	\citet{Mukhopadhyay2017} developed a site-type-specific regression on the AQUM data using our same classification for the site type. The key feature of their model is the specification of a non-stationary spatio-temporal process, which leads to a better predictive performance compared to the stationary Gaussian process (GP) in their application. 
	
	To obtain a like-for-like comparison, we also include a site-type-specific regression on the PCM data and implement both the stationary and the non-stationary versions of this model. 
	
	Both AQUM ($\boldsymbol{X_1}$) and PCM ($\boldsymbol{X_2}$) are interpolated at the monitoring site locations through bilinear interpolation. 
	
	The hierarchical model specification in this case is: 
	
	$y(\boldsymbol{s},t) \sim N(\eta(\boldsymbol{s},t), \sigma^{2}_{\epsilon}) $
	
	$\eta(\boldsymbol{s},t) = \mu(\boldsymbol{s},t) + \nu(\boldsymbol{s},t)$
	
	with 
	
	$\mu(\boldsymbol{s},t) = \gamma_0 + \gamma_1X_1(\boldsymbol{s},t) + \sum_{k=1}^2{\delta_k(\boldsymbol{s})(\gamma_{0k}+\gamma_{1k}X_1(\boldsymbol{s},t))}$ 
	
	for the model with AQUM only, and 
	
	$\mu(\boldsymbol{s},t) = \gamma_0 + \gamma_1X_1(\boldsymbol{s},t) + \gamma_2X_2(\boldsymbol{s},t) + \sum_{k=1}^2{\delta_k(\boldsymbol{s})(\gamma_{0k}+\gamma_{1k}X_1(\boldsymbol{s},t)+\gamma_{2k}X_2(\boldsymbol{s},t))}$ 
	
	for the model with AQUM and PCM.
	
	Here $k=0$ indicates rural site type (baseline), $k=1$ urban and $k=2$ road-kerbside, $\delta_k(\boldsymbol{s})$ is an indicator function equal to 1 if site $\boldsymbol{s}$ is of type $k$ and 0 otherwise, $\gamma_0$, $\gamma_1$ and $\gamma_2$ are the baseline intercept and slopes for $\boldsymbol{X_1}$ and $\boldsymbol{X_2}$, while $\gamma_{0k}$, $\gamma_{1k}$ and $\gamma_{2k}$ are site-type-specific adjustments to the baseline intercept and slopes.
	
	For the spatio-temporal process $\boldsymbol{\nu}$ we first assume a stationary time-independent GP with zero mean and exponential correlation function (note that $\nu_t(\boldsymbol{s})=\nu(\boldsymbol{s},t)$):
	
	$\boldsymbol{\nu}_t \sim N(\boldsymbol{0}, \sigma^2_\nu\boldsymbol{H}_\nu(\phi))$ , where $\boldsymbol{H}_\nu(\phi) = corr(\nu_t(\boldsymbol{s}),\nu_t(\boldsymbol{s}')) = exp(-||s-s'||\phi)$.
	
	Then we specify a non-stationary covariance structure as in \citet{Sahu2015a}: given a GP $\boldsymbol{\nu}_t^*$ defined on a set of $m=25$ knot locations $\boldsymbol{\nu}_t^* \sim MVN(\boldsymbol{0}, \sigma^2_\nu \boldsymbol{H}_{\nu^*}(\phi))$, the Gaussian predictive process (GPP) at a new location $\boldsymbol{s}$ is defined as  $\tilde{\nu}_t(\boldsymbol{s}) = E[\nu_t(\boldsymbol{s})|\boldsymbol{\nu}_t^*]$. From multivariate Gaussian theory it follows that $\tilde{\boldsymbol{\nu}}_t=C^* \boldsymbol{H}^{-1}_{\nu^*}(\phi)\boldsymbol{\nu^*}_t$ with $C^*$ being the cross-correlation function between $\boldsymbol{\nu}_t$ and $\boldsymbol{\nu^*}_t$. 
	
	The non-stationarity and the anisotropy are given by the fact that $corr(\tilde{\nu}_t(\boldsymbol{s}),\tilde{\nu}_t(\boldsymbol{s'}))=\boldsymbol{c}^*(\boldsymbol{s})^T\boldsymbol{H}^{-1}_{\nu^*}(\phi)\boldsymbol{c}^*(\boldsymbol{s'})$ which depends on both $\boldsymbol{s}$ and $\boldsymbol{s'}$ and not only on the separation vector or the distance between locations.
	
	To introduce temporal dependence we specify a first order autoregressive model for $\boldsymbol{\nu}_t^*$.
	
	The choice of knot locations, model specification and prior distributions is based on \citet{Mukhopadhyay2017}. This is justified by the fact that we use the same data sources on a subregion of their study area.

	\subsection{Validation and predictive capability measures} \label{s:valid}
	
	We compared the predictive capability through proper scoring rules \citep{Gneiting2007} such as the cross-validated logarithmic score (logScore), the Continuous Ranked Probability Score (CRPS) and the Root Mean Squared Error (RMSE), and also the Predictive Model Choice Criterion (PMCC) proposed by \citet{Gelfand1998}.
	
	Furthermore, we reported  the correlation between the observed and the predicted values for the validation sites (COR), the Mean Absolute Percentage Error (MAPE), and the 95\% coverage (COV), defined as the percentage of times that the observed value falls within the 95\% credibility interval of the sampled posterior marginal.
	
	To measure the predictive capability we need to report the fitted values at the validation sites on the scale of the outcome, as cannot compare the observed values with the fitted values because they are not accounting for the measurement error, but only for the uncertainty associated with the model parameters. In order to do so, we draw 50 values from the marginal posterior of the measurement error $p(\sigma^2_{\epsilon_3} |\boldsymbol{y_3})$ first, then draw $\boldsymbol{\eta_3}$ from its conditional posterior $p(\boldsymbol{\eta_3} | \sigma^2_{\epsilon_3}, \boldsymbol{y_3})$ using the simulated values of $\sigma^2_{\epsilon_3}$ \citep{BDA}. These values are used as mean and variance of a Normal distribution, from which we sampled values at each site (sample size = 100). 
	
	The following are the formulas for the different measures of predictive capability used in the paper. For simplicity we apply a slight change of notation here: $y_{jt}$ indicates the observed value at monitor $j$ ($m$ is the number of validation monitors) and day $t$ ($t=1,\dots,T$, $T$=1826), and $\hat{y}_{jt}$ is the corresponding predicted value obtained as mean of the vector of $Q=100\times50$ sampled values $\boldsymbol{\hat{y}}_{jt} = \hat{y}_1, \dots, \hat{y}_Q \sim F_{jt}$, $F_{jt}$ being the empirical distribution function of $\boldsymbol{\hat{y}}_{jt}$. 
		
	\begin{align*} 
	RMSE &=\sqrt{\frac{1}{mT} \sum_{j=1}^{m}\sum_{t=1}^{T}(y_{jt}-\hat{y}_{jt})^2}\\
	MAPE &=\frac{1}{mT} \sum_{j=1}^{m}\sum_{t=1}^{T}\frac{|y_{jt}-\hat{y}_{jt}|}{y_{jt}} \cdot 100 \\
	PMCC &= \sum_{j=1}^{m}\sum_{t=1}^{T}(y_{jt}-\hat{y}_{jt})^2 +   \sum_{j=1}^{m}\sum_{t=1}^{T}VAR(\boldsymbol{\hat{y}}_{jt}) \\
	CRPS &= \frac{1}{mT} \sum_{j=1}^{m}\sum_{t=1}^{T}CRPS(F_{jt},y_{jt})  \quad \text{, with} \\
	CRPS(F_{jt},y_{jt}) &=\frac{1}{Q}  \sum_{q=1}^{Q}|\hat{y}_q-y_{jt}| - \frac{1}{2Q^2}\sum_{q=1}^{Q}\sum_{r=1}^{Q}|\hat{y}_q-\hat{y}_r|  
	\end{align*}
	
	For each model under comparison, the predictive capability measures presented above are computed pooling together the 6 validation sets. It can be calculated by day, by site, by site type or across all sites to obtain specific and global measures, with lowest measures indicating the best predictive performance.

	\subsection{Predictions on a regular grid} \label{s:predictions}
	
	From the joint model, we extract daily predictions of NO$_2$ concentration on a regular grid that covers the study area. For the grid we choose an intermediate spatial resolution between PCM and AQUM data to limit the computational burden while retaining spatial variability.
	
	In order to provide the predictions in a reasonable time, we extract samples from the joint posterior marginals and estimate the linear predictor at each time-location for the 1826 days on the regular grid \citep{thomas2019}.
	
	We compute the predictions from the model that includes all monitors as training set.
		
	We extract samples from the posterior marginals of the model components in order to reconstruct the linear predictor at each time-location for the 1826 days on the regular grid, as:
	\begin{equation*} 
		\eta_3(\boldsymbol{s},t) = \alpha_3 + \beta_{k_s} + \lambda_{1,3} z_1(\boldsymbol{s}) +   \lambda_{2,3} z_2(t) + z_3(t,k{_s})  
	\end{equation*}
	In particular, following the tutorial by \citet{Bakka2017}, we obtain samples from the posterior of the intercept $\alpha_3$ and $\beta_{k_s}$ for each site type, samples from the posterior of $\lambda_{1,3}\boldsymbol{z_1}$ at the mesh nodes and reproject it on the prediction grid, samples from the posterior of $\lambda_{2,3}\boldsymbol{z_2}$ at each time point (days), and samples from the posterior of $\boldsymbol{z_3}$ for each day and site type.
		
	Note that in order to predict at the grid locations we need to know the value of site type classification for each grid point. With this aim we built a function which assigns each location to road-kerb side, urban or rural depending on the distance from any road as well as using the Corine land cover for the year 2012 for the UK, Jersey and Guernsey shapefile from the Centre for Ecology and Hydrology \citep{ceh}. See Appendix \ref{a:sitetype} for more details. 
		
	For each sample we then sum up the samples from the fixed effects and random effects to reconstruct the linear predictor, then the prediction is given by average across all samples.

	\section{Results} \label{s:results}
	
	In this section we present the results of the model comparison, with particular focus on the advantages of the proposed joint model, and the daily predictions that we obtained from the best model.

	\subsection{Model comparison} \label{s:mod_comp}
	
	In order to show whether the inclusion of more than one extra data source actually improves the model predictive capability, we compare our proposed joint model with the corresponding models that include only AQUM or PCM. 
	
	For AQUM we assume a spatio-temporal effect or temporal-only effect when PCM is included. 
	
	We also compare our joint model with other well established data integration techniques, the simple interpolation models described in section \ref{s:bilint} and the more complex ones described in section \ref{s:sujit}. 
	
	Besides providing information about all the sources of uncertainty, all the joint models have better performance than the models where the misaligned data are interpolated, even allowing for non-stationarity (see table \ref{t:mod_comp}). 
	
	However, the AQUM data do not seem to provide much information, in fact the model that includes only AQUM has a far worse performance than the one only including PCM. In addition, allowing for a spatial effect on AQUM does not improve the prediction, for the model where PCM is also included. 
	This can be explained by the fact that the time-sitetype interaction $\boldsymbol{z_3}$ replaces the role of AQUM in capturing the temporal trend when we remove AQUM from the model and the temporal information is still provided by the numerous monitoring stations, while there is no other structured spatial component that compensates for PCM when it is removed. 
	
	Furthermore, as we focus here on spatial prediction rather than temporal forecasting, removing AQUM is less of a burden on the model performance in terms of predictive capability.
	
	Nevertheless the model including both AQUM and PCM has the best performance in terms of PMCC and CRPS and we will report the results from this model in the next section. 
	
	Note that the predictive capability measures of the models in section \ref{s:sujit} cannot be compared with the others due to the different model structure. Only the PMCC and the 95\% coverage are comparable and reported in table \ref{t:mod_comp}.
	
	With regard to these models, allowing for non-stationarity and anisotropy leads to very little gain compared to the introduction of an additional source of data at high spatial resolution. In general, their performance is almost as poor as having a linear effect on interpolated covariates.  
	
	\begin{table}
		\centering
		\begin{threeparttable}[b]
			\small{
				\caption{Model comparison in terms of predictive capability}
				\label{t:mod_comp}
				\begin{tabular}{p{0.2cm}p{0.1cm}p{0.6cm}p{0.6cm}p{0.6cm}p{0.6cm}p{0.6cm}p{0.6cm}p{0.6cm}p{0.6cm}}
					\noalign{\smallskip}\hline\noalign{\smallskip}                                                                                                                          
					&& \multicolumn{6}{c}{Predictive capability}  \\		
					\cline{ 3-8}\\ [-6pt]
					\hline\noalign{\smallskip} 
					Model\tnote{**}     						 		
					&& \multicolumn{1}{c}{PMCC} 
					&  \multicolumn{1}{c}{CRPS}
					&  \multicolumn{1}{c}{RMSE} 
					&  \multicolumn{1}{c}{MAPE}
					&  \multicolumn{1}{c}{CORR}
					& \multicolumn{1}{c}{COV}  \\
					\noalign{\smallskip}\hline\noalign{\smallskip} 				
					\multicolumn{1}{r}{$AQUM(s,t)$ joint}   && 18277 & 0.0523 & 0.5725 & 16.71\% & 65.83\% & 78.15\%  \\
					\multicolumn{1}{r}{$PCM(s)$ joint}   && 14018 & 0.0372 & 0.4615 & 13.54\% & 76.77\% & 86.87\%  \\
					\multicolumn{1}{r}{$\boldsymbol{AQUM(s,t)+PCM(s)}$ \bf{joint}}   && \bf{13621} & \bf{0.0338} & 0.4665 & 13.67\%  & 76.08\%  &  84.66\%    \\
					\hline
					\multicolumn{1}{r}{$AQUM+PCM$ bilinear interpolation}  	&& 82970 & 0.2560 &  0.6911 & 17.57\% & 67.14\% & 68.55\%  \\
					\multicolumn{1}{r}{$AQUM+PCM$ kriging estimates}  	&& 35017 & 0.2220 & 0.4964 & 14.58\% & 73.13\% & 75.27\%  \\
					\hline
					\multicolumn{1}{r}{\begin{tabular}[r]{@{}r@{}}$AQUM$, non-stationary\\\citep{Mukhopadhyay2017} \end{tabular}}  	&&   79542\tnote{*}  &  &  &  &  & 60.74\%  \\
					\multicolumn{1}{r}{\begin{tabular}[r]{@{}r@{}}$AQUM+PCM$, non-stationary\\\citep{Mukhopadhyay2017} \end{tabular}}  	&&   75506\tnote{*}  &  &  &  &  & 62.13\%  \\
					\multicolumn{1}{r}{\begin{tabular}[r]{@{}r@{}}$AQUM+PCM$, stationary\\ \citep{Mukhopadhyay2017} \end{tabular}}  	&&   75510\tnote{*}  &  &  &  &  & 62.13\%  \\
					\noalign{\smallskip}\hline\noalign{\smallskip}
			\end{tabular}}
			\begin{tablenotes}
				\item[*]As provided by spT.Gibbs function in R package spAir.
				\item[**] $(s)$ indicates spatial-only random effects; $(t)$ indicates temporal-only random effects; $(s,t)$ indicates additive spatial and temporal random effects. When not specified, a linear effect is assumed as described in section \ref{s:competitors}. 
			\end{tablenotes}
		\end{threeparttable}
	\end{table}

	\subsection{Results from the complete joint model}			
	
	We report the results for the joint model that includes spatial and temporal effects on AQUM and spatial effect on PCM, re-ran using all monitors as training data. 
	
	Looking at the summary reported in table \ref{t:hyperpar} we see that, as expected, there is an increase in the NO$_2$ concentration going from rural to road-kerb side locations, but not for urban. For the spatial latent field $\boldsymbol{z_1}$, the estimated empirical range, i.e. the distance after which the spatial correlation function drops to 0.13 \citep{Lindgren2015}, is 177 Km, corresponding to approximately 50\% of the maximum extension of the spatial domain. 	
	
		\begin{table}
			\centering
			\begin{threeparttable}[b]
				\caption{Summary of model parameters and hyperparameters}
				\label{t:hyperpar}
				\begin{tabular}{rrrrrr}
					\hline\noalign{\smallskip}  
					\multicolumn{1}{l}{} & \multicolumn{1}{c}{mean} & \multicolumn{1}{c}{SD} & \multicolumn{1}{c}{0.025q} & \multicolumn{1}{c}{median} & \multicolumn{1}{c}{0.975q} \\
					\noalign{\smallskip}\hline\noalign{\smallskip}           
					$\alpha_1$		&  2.0653  & 0.0308  &   2.0048 &  2.0653  &   2.1258 \\
					$\alpha_2$		&  2.5793  & 0.0259  &   2.5285 &  2.5793  &   2.6301 \\
					$\alpha_3$		&  2.4722  & 0.0236  &   2.4258 &  2.4722  &   2.5186 \\
					$\beta_{URB}$	&  -0.1716 & 0.0047  &  -0.1808 & -0.1716  &  -0.1624 \\ 
					$\beta_{RKS}$	&  0.3764  & 0.0047  &   0.3673 &  0.3764  &   0.3856 \\
					$\sigma^2_{\epsilon_1}$	  	&    0.0003 &  0.0000  &     0.0003 &     0.0003  &     0.0003  \\
					$\sigma^2_{\epsilon_2}$	  	&    0.0303 &  0.0000  &     0.0303 &     0.0303  &     0.0303  \\
					$\sigma^2_{\epsilon_3}$	  	&    0.0213 &  0.0000  &     0.0213 &     0.0213  &     0.0213  \\
					$\sigma^2_{z_1}$	  		&    2.0729 &  0.0385  &     1.9815 &     2.0803  &     2.1225  \\
					$\sigma^2_{z_2}$	  		& 2626.3 &  7.960  &  2607.8 &  2627.7  &  2637.7  \\
					$\sigma^2_{z_3}$	  		&    0.0013 &  0.0000  &     0.0013 &     0.0013  &     0.0013  \\
					$r_{z_1}$ (km)       & 177.8 &  0.227   & 177.2  &  177.77  &  256.0  \\ 
					$\rho_{z_3}$		 & 0.5702   &  0.0005   &   0.5689  &    0.5700  &    0.6869  \\ 
					$\lambda_{1,2}$      & 1.1000   &  0.0001   &   1.0996  &    1.0999  &    1.1345  \\ 
					$\lambda_{1,3}$      & 1.2999   &  0.0003   &   1.2989  &    1.2998  &    1.3990  \\ 
					$\lambda_{2,3}$      & 0.8995   &  0.0004   &   0.8977  &    0.8995  &    0.9003  \\ 
					\noalign{\smallskip}\hline\noalign{\smallskip}
				\end{tabular}	
			\end{threeparttable}		
		\end{table}

	The scaling parameters $\lambda_{i,j}$ are all different from 1, meaning the spatial field for PCM needs to be rescaled for AQUM ($\lambda_{1,2}=1.1$) and for the monitor observations ($\lambda_{1,3}=1.3$), and the temporal latent field for AQUM is also calibrated against the monitor observations with $\lambda_{2,3}=0.9$.
	
	The intercepts $\alpha_i$ represent the overall mean of PCM, AQUM and ground observations respectively. 
	
\begin{figure}
		\centerline{\includegraphics[width=.8\textwidth]{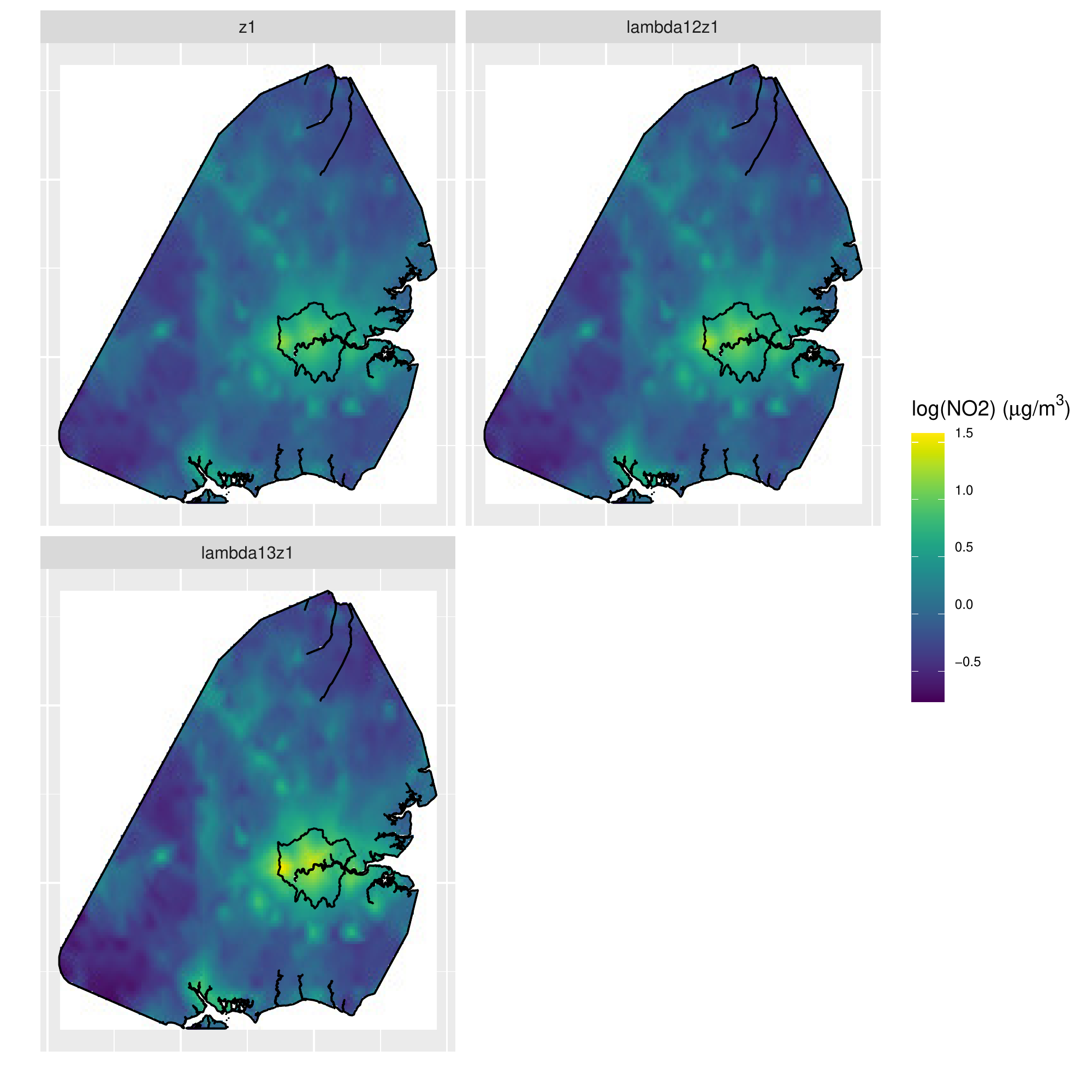}}
		\caption{Posterior mean of the latent spatial field $\boldsymbol{z_1}$ and of the rescaled spatial fields $\lambda_{1,2}\boldsymbol{z_1}$ and $\lambda_{1,3}\boldsymbol{z_1}$.}
		\label{f:z1}
	\end{figure}

	The spatial latent field $\boldsymbol{z_1}$ (Fig.~\ref{f:z1}) shared between the PCM data, the AQUM data and the monitor observations shows the traffic-driven characteristics of NO$_2$ as we can recognize higher values in correspondence of motorways and major city centers. The rescaled fields are reported as well and for $\lambda_{1,3}\boldsymbol{z_1}$ the magnifying effect of the scaling parameter $\lambda_{1,3}=1.3$ is particularly visible.
	
\begin{figure}
		\centerline{\includegraphics[height=\textwidth,angle =-90]{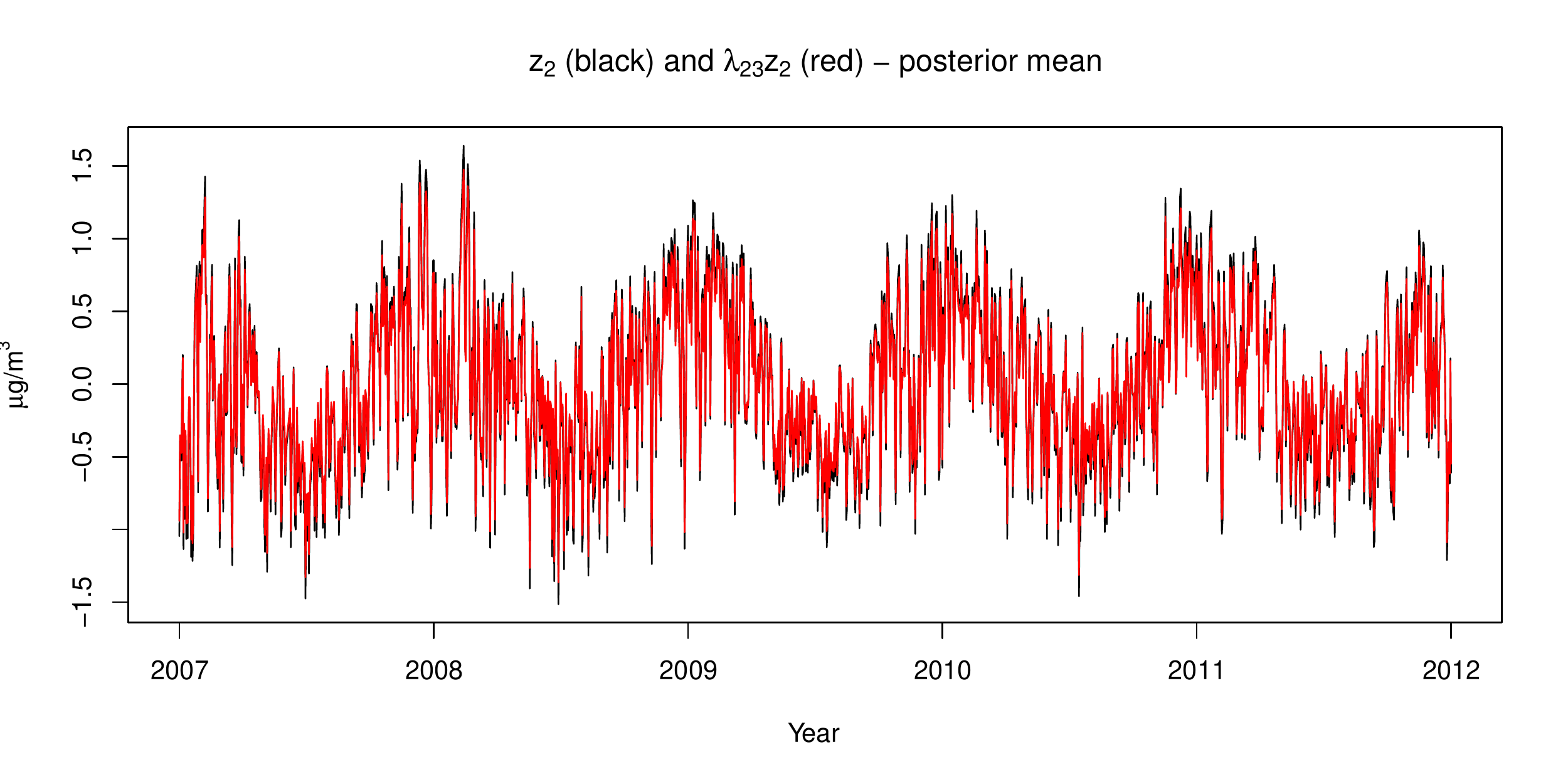}}
		\caption{Posterior mean of the latent temporal field $\boldsymbol{z_2}$ (black line) and of the rescaled temporal field $\lambda_{2,3}\boldsymbol{z_2}$ (red line).}
		\label{f:z2}
	\end{figure}

	Figure \ref{f:z2} shows the temporal latent field $\boldsymbol{z_2}$ shared between the AQUM data and the monitor observations which captures the seasonality of NO$_2$, and the rescaled field $\lambda_{2,3}\boldsymbol{z_2}$ which is shrinked by the scaling parameter $\lambda_{2,3}=0.9$.
	
	The latent fields $\boldsymbol{z_1}$ and $\boldsymbol{z_2}$ are both centred in zero as the large scale component of PCM and AQUM is captured by their intercepts $\alpha_1$ and $\alpha_2$.
	
\begin{figure}
		\centerline{\includegraphics[height=\textwidth,angle =-90]{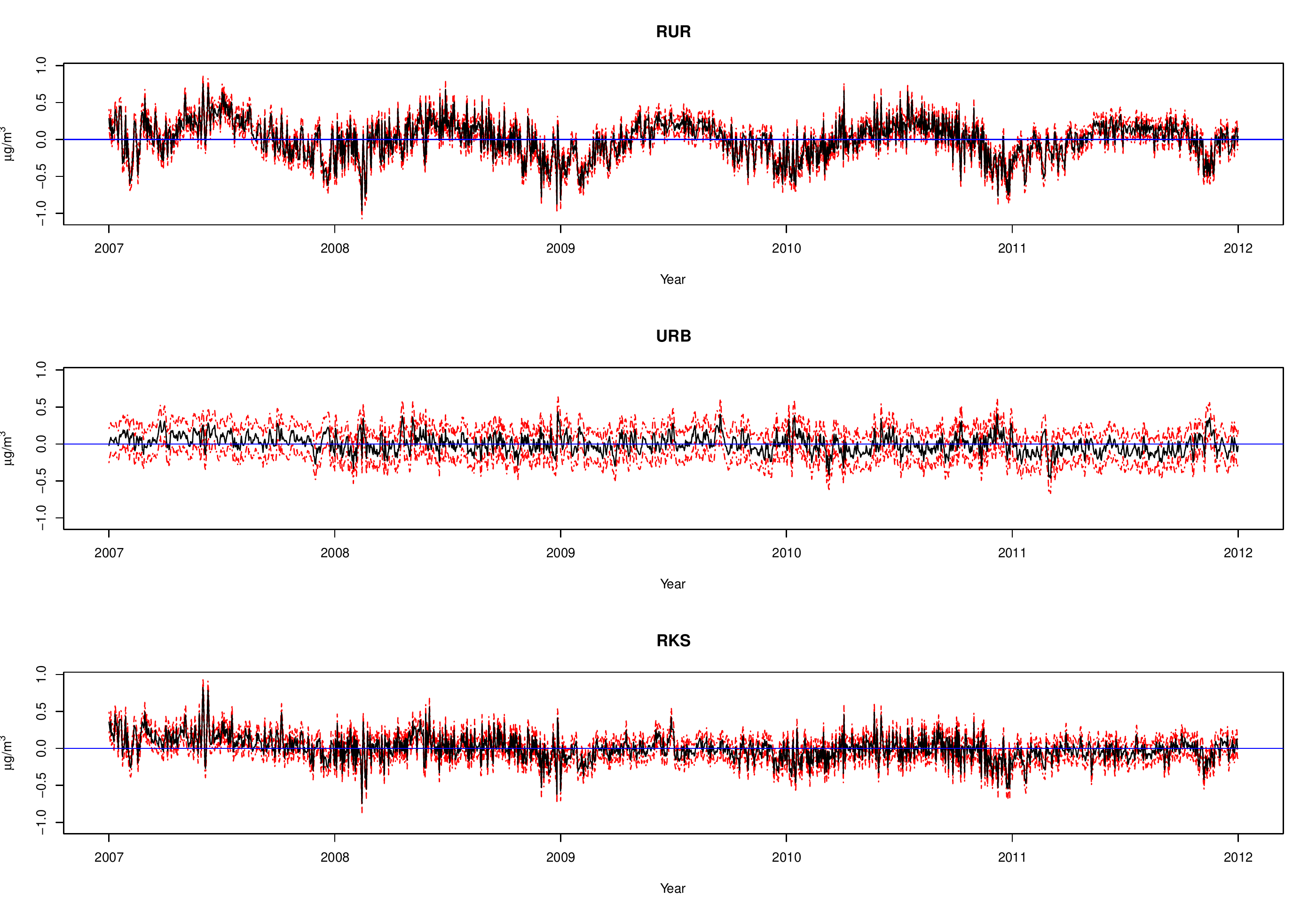}}
		\caption{Time-sitetype interaction $\boldsymbol{z_3}$. Posterior mean in red, 95\% CI in black dashed lines.}
		\label{f:time_sitetype}
	\end{figure}

	Finally, the time-sitetype interaction $\boldsymbol{z_3}$ in Fig.~\ref{f:time_sitetype} shows that there is some residual site-type-specific temporal variability, especially for urban and road-kerb side monitors, which is not captured by the main temporal component $\boldsymbol{z_2}$. 
	
	\subsubsection{Daily predictions} \label{s:daily_predictions}
	
\begin{figure}
		\centerline{\includegraphics[width=\textwidth]{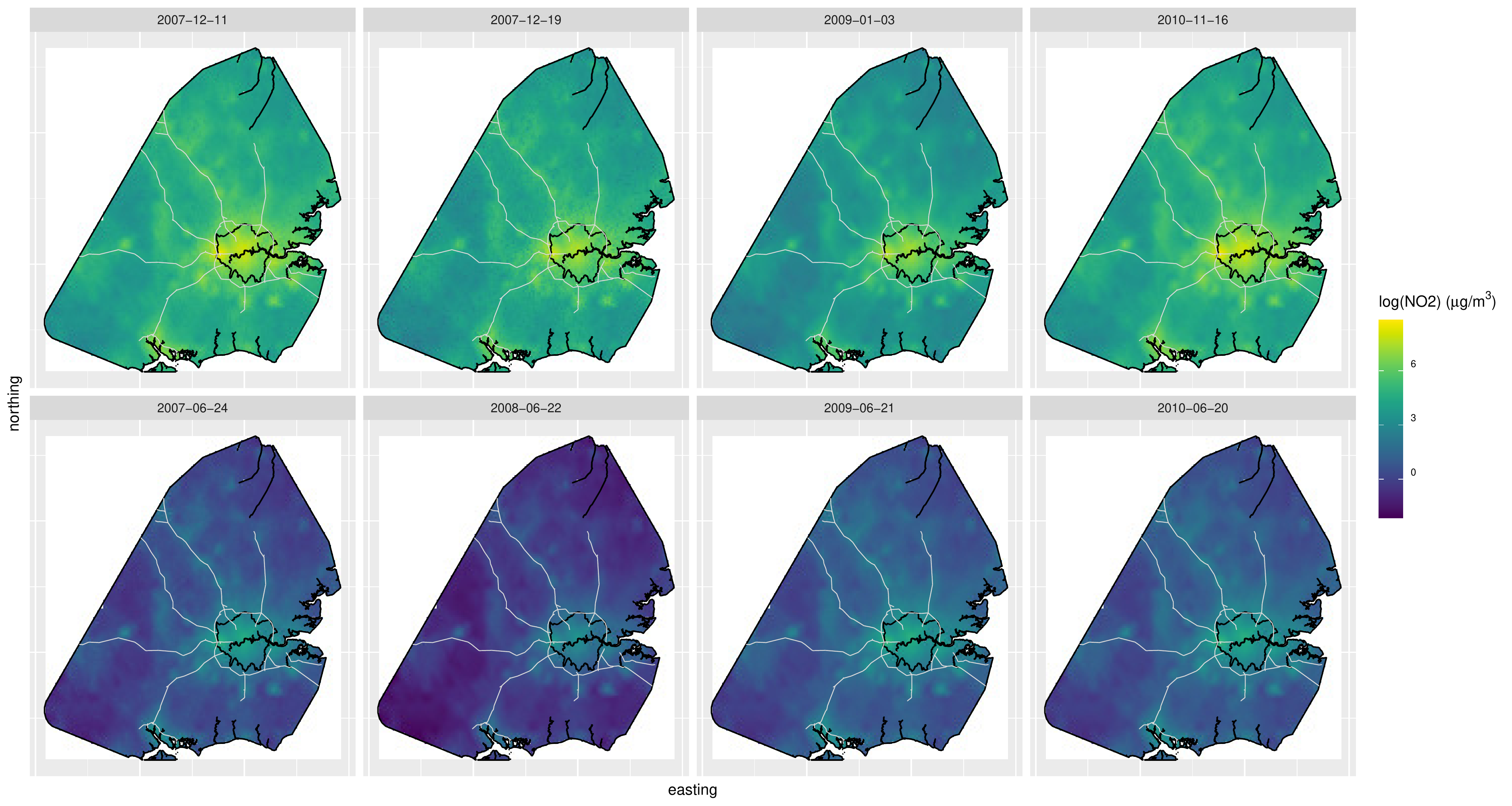}}
		\caption{Daily predictions for four days in which an air pollution event was registered (top row) and four days with reported low air pollution concentration (bottom row).}
		\label{f:daily_pred}
	\end{figure}

	We selected four NO$_2$ pollution episodes reported by the LondonAir website \citep{londonair} and compared the predictions for these four days with four randomly selected summer Sundays across the study period, where we expect to see low levels of NO$_2$. The predictions show the expected behaviour, with high predicted concentrations during the pollution episodes and low concentrations during the selected Sundays (Fig.~\ref{f:daily_pred}). 
	
	A layer with the roads classified as motorways is plotted on top of each map, showing correspondence between the highest predicted levels of NO$_2$ and the major roads. This is expected because NO$_2$ is a highly traffic-driven pollutant. A peak of NO$_2$ concentration can also be observed in the area of Heathrow airport, on the left of Greater London, which is characterised by the highest levels also on low concentration days.

	\section{Conclusion and discussion} \label{s:conclusion}
	
	We implemented a hierarchical Bayesian model to estimate air pollution concentration, combining misaligned data sources with a joint approach. This approach can be considered in between Bayesian melding and calibration, and it is the first attempt at implementing such methods on spatio-temporal air pollution data in \texttt{R-INLA}.

	The proposed model includes information on the site type as well as output from two different numerical models characterised by spatial and temporal variability and accounting for traffic, chemistry, land use and meteorological covariates. Our method is transferable to any available data sources, however the interpretation of the results may change according to their intrinsic characteristics, in particular referring to the information included in the deterministic or LUR models. 
	
	We show that including more than one covariate at different spatial and temporal resolution increases model predictive capability. However, removing AQUM has proven not to be detrimental, but this could be justified with the fact that we are not doing temporal forecasting. 
	
	Overall we prove that using as much spatial and temporal information as possible is more beneficial than increasing the complexity of the random effect structure.
	
	A time-site type interaction was added to the model to account for residual temporal variability observed when looking at the site type-specific residuals. 
	
	The advantages of our method are manyfolds: first, reconstructing the entire latent field in a Bayesian approach provides us with the marginal posterior distribution for all the uncertainty parameters, allowing us to correctly quantify the uncertainty associated with our predictions and the deterministic models, that is not possible to obtain with other downscalers and non-model-based solutions; second, unlike the spatio-temporal downscaler proposed by \citet{Berrocal2012}, our model reconstructs the latent fields of the misaligned covariates as a whole, rather than locally. For the same reason, in order to obtain daily predictions at new locations there is no need to calculate the value of the misaligned covariates at the prediction locations, as the model already estimates the whole latent field.

	Our analysis presents some limitations related on one side to the computational requirements of INLA due to the high number of parameters, and on the other side to the generalizability of the results, as the models are quite data-sensitive. In particular, we have very few rural sites even though we extended the study domain outside Greater London, suggesting the presence of preferential sampling that we did not account for. Furthermore, we made assumptions of stationarity and isotropy which may not hold when extending the spatial domain to bigger areas.
	
	As a next step we will extend the joint model to a multivariate version including other pollutants, such as PM$_{10}$ or O$_{3}$. 
	
	In the future, the predicted air pollution concentration with associated measure of uncertainty could be used as exposure in an epidemiological model, allowing for uncertainty propagation.

\section*{Acknowledgements}
The authors wish to thank the King's College London Environmental Research Group and Paul Agnew from the Met Office for providing the data, Ben Barratt (King's College London) and Monica Pirani (Imperial College London) for the valuable comments, and Haakon Bakka for the priceless explanations on INLA functions. This work was supported by the Medical Research Council [grant number MR/M025195/1]. Chiara Forlani was funded by Imperial College London President's PhD scholarship. Michela Cameletti has been supported by the PRIN EphaStat Project (Project No. 20154X8K23, \url{https://sites.google.com/site/ephastat/}) provided by the Italian Ministry for Education, University and Research.

\bibliographystyle{wb_env}  
\bibliography{library}  

\appendix
\counterwithin{figure}{section}
\counterwithin{table}{section}

\section{Further figures} \label{a:figures}

Additional figures mentioned in the paper are reported here. 

\begin{figure}
	\centerline{\includegraphics[height=\textwidth, angle =-90]{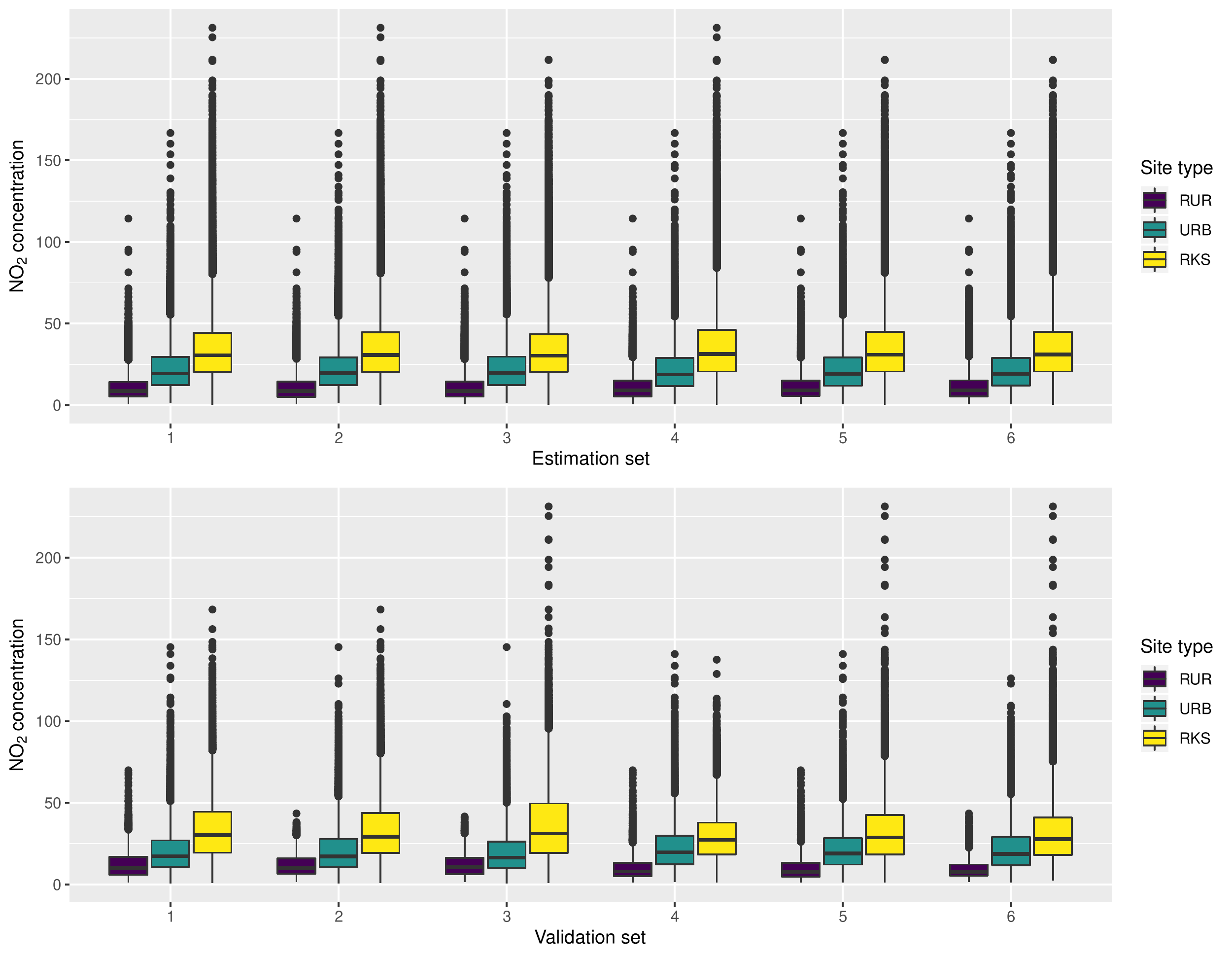}}
	\caption{Boxplot of NO$_2$ concentration by site type for each estimation and validation set.}
	\label{f:descriptive}
\end{figure}

\begin{figure}
	\centerline{\includegraphics[width=.95\textwidth]{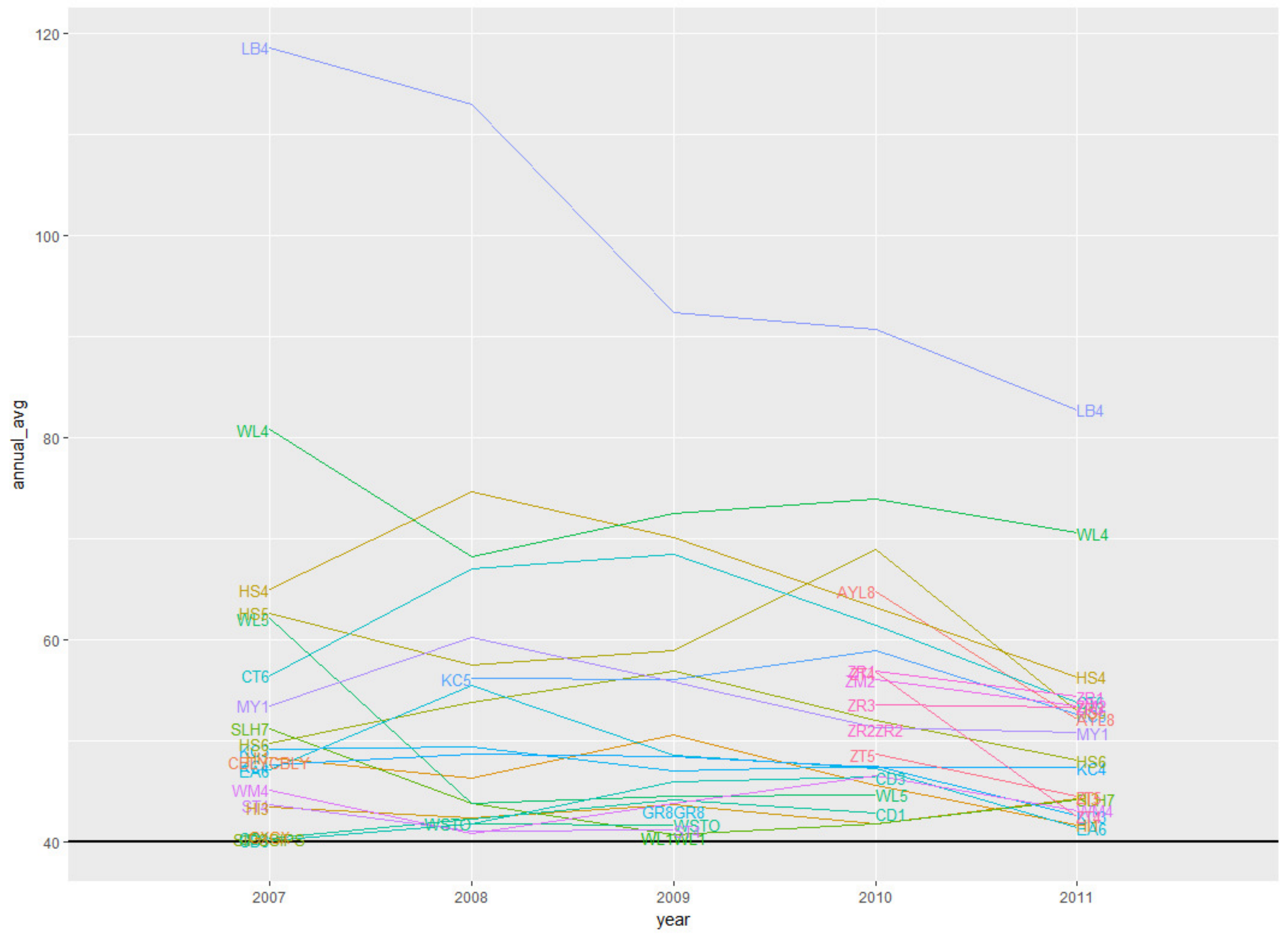}}
	\caption{Annual averages of NO$_2$ hourly concentration for monitoring sites that exceeded the WHO annual threshold of 40$\mu g/m^3$ at least once in the study period.}
	\label{f:exceedance}
\end{figure}

\section{Comparison of separate models for AQUM and PCM data} \label{a:separate_models}

Table \ref{t:sep_mod} shows the deviance information criterion (DIC) and the logarithmic score for the models implemented separately for AQUM and PCM.

\begin{table} 
	\centering
	\begin{threeparttable}[b]
		\caption{Comparison of performance for separate models of AQUM and PCM}
		\label{t:sep_mod}
		\centering
		\begin{tabular*}{.5\textwidth}{@{}l@{\extracolsep{\fill}}c@{\extracolsep{\fill}}c@{}}
			\hline 							
			Model \tnote{*} & DIC 	& logScore		 \\
			\hline
			(i) $AQUM(t)$ & 1155407 &  0.64	\\
			(ii) $\boldsymbol{AQUM(s+t)}$  	& \bf{659641} & \bf{0.36} 	\\
			(iii) $AQUM(s*t)$ 	& 1418821 &	0.79 \\
			(i) $\boldsymbol{PCM(s)}$  	& \bf{-373298}  &	\bf{-0.85}  \\
			(ii) $PCM(s+t)$  	& -403560 & -0.91	\\
			(iii) $PCM(s*t)$ 	&  -456714 & -1.04\\
			\hline
		\end{tabular*}
		\begin{tablenotes}
			\item[*] $(t)$ indicates temporal-only random effect; \\
			$(s)$ indicates spatial-only random effect; \\
			$(s+t)$ indicates additive spatial and temporal effects; \\
			$(s*t)$ indicates space-time interaction. 
		\end{tablenotes}
	\end{threeparttable}
\end{table}

We need to include a spatial component for AQUM as this significantly improves the model performance. 

According to the reported measures, the spatial-only model for PCM is the worst in terms of goodness of fit. However, the temporal component seems negligible and when including a space-time interaction the range becomes unreasonable (5230 Km). For these reasons, and to limit the computational burden, we decided to keep only the spatial component when modelling PCM, also because the temporal information is provided by the monitors and the AQUM data at a much higher resolution (daily instead of annual).

\section{INLA-SPDE} \label{a:inla}

In this section we define the class of models on which we can perform Bayesian inference using INLA and briefly introduce how the inference is computed, following the notation in \citet{Blangiardo2015}.

Let us consider a set of data $\boldsymbol{y}=(y_1,\dots,y_n)$ with distribution characterized by a parameter $\boldsymbol{\mu}$, usually the mean, defined as a function of an additive linear predictor $g(\boldsymbol{\mu}) = \boldsymbol{\eta}$:
\begin{equation*}
g(\mu_i) = \eta_i = \alpha + \sum_{m=1}^M \beta_mx_{mi} + \sum_{l=1}^L f_l(z_{li})
\end{equation*}

Defining the vector of parameters $\boldsymbol{\theta}=(\alpha,\boldsymbol{\beta},\boldsymbol{f})^T$ and the vector of hyperparameters $\boldsymbol{\psi}=(\psi_1,\dots,\psi_K)$, the likelihood is given by
\begin{equation*}
p(\boldsymbol{y}|\boldsymbol{\theta},\boldsymbol{\psi}) = \prod_{i=1}^np(y_i|\theta_i,\boldsymbol{\psi})
\end{equation*}

We assume the latent field $\boldsymbol{\theta}$ to be multivariate Normal with precision matrix $\boldsymbol{Q}$ and conditionally independent, i.e. a Gaussian Markov Random Field (GMRF): 
\begin{equation*}
\boldsymbol{\theta} \sim MVN(\boldsymbol{0}, \boldsymbol{Q}^{-1}(\boldsymbol{\psi}))
\end{equation*}

The Markov property ensures the sparsity of the precision matrix $\boldsymbol{Q}$. 

The aim of Bayesian inference is to obtain the posterior marginal distributions for all the model parameters $p(\theta_i|\boldsymbol{y}) = \int p(\theta_i,\boldsymbol{\psi}|\boldsymbol{y}) d\boldsymbol{\psi} = \int p(\theta_i|\boldsymbol{\psi},\boldsymbol{y})p(\boldsymbol{\psi}|\boldsymbol{y}) d\boldsymbol{\psi}$
and hyperparameters $p(\psi_k|\boldsymbol{y}) = \int p(\boldsymbol{\psi}|\boldsymbol{y}) d\boldsymbol{\psi}_{-k}$.

Therefore we first need to compute (i) the joint posterior marginal of the hyperparameters $p(\boldsymbol{\psi}|\boldsymbol{y})$ and (ii) the posterior conditional distributions $p(\theta_i|\boldsymbol{\psi},\boldsymbol{y})$.

Within such class of models, which includes a wide range of possible model specifications, we can compute these distributions through a Laplace approximation: 
\begin{align*}
p(\boldsymbol{\psi}|\boldsymbol{y}) &= \frac{p(\boldsymbol{\theta},\boldsymbol{\psi}|\boldsymbol{y})}{p(\boldsymbol{\theta}|\boldsymbol{\psi},\boldsymbol{y})} 
= \frac{p(\boldsymbol{y}|\boldsymbol{\theta},\boldsymbol{\psi})p(\boldsymbol{\theta},\boldsymbol{\psi})}{p(\boldsymbol{y})}\frac{1}{p(\boldsymbol{\theta}|\boldsymbol{\psi},\boldsymbol{y})} 
\propto \frac{p(\boldsymbol{y}|\boldsymbol{\theta},\boldsymbol{\psi})p(\boldsymbol{\theta},\boldsymbol{\psi})}{p(\boldsymbol{\theta}|\boldsymbol{\psi},\boldsymbol{y})} \\
&\approx \frac{p(\boldsymbol{y}|\boldsymbol{\theta},\boldsymbol{\psi})p(\boldsymbol{\theta},\boldsymbol{\psi})}{\tilde{p}(\boldsymbol{\theta}|\boldsymbol{\psi},\boldsymbol{y})}\bigg|_{\boldsymbol{\theta=\theta^*(\boldsymbol{\psi})}} =: \tilde{p}(\boldsymbol{\psi}|\boldsymbol{y})
\end{align*}

where $\tilde{p}(\boldsymbol{\theta}|\boldsymbol{\psi},\boldsymbol{y})$ is the Gaussian Laplace approximation of $p(\boldsymbol{\theta}|\boldsymbol{\psi},\boldsymbol{y})$ around its mode $\theta^*(\boldsymbol{\psi})$.

For $p(\theta_i|\boldsymbol{\psi},\boldsymbol{y})$ we consider a partition $\boldsymbol{\theta}=(\theta_i,\boldsymbol{\theta}_{-i})$:
\begin{align*}
p(\theta_i|\boldsymbol{\psi},\boldsymbol{y}) &= \frac{p((\theta_i,\boldsymbol{\theta}_{-i})|\boldsymbol{\psi},\boldsymbol{y})}{p(\boldsymbol{\theta}_{-i}|\theta_i,\boldsymbol{\psi},\boldsymbol{y})}
= \frac{p(\boldsymbol{\theta},\boldsymbol{\psi}|\boldsymbol{y})}{p(\boldsymbol{\psi}|\boldsymbol{y})} \frac{1}{p(\boldsymbol{\theta}_{-i}|theta_i,\boldsymbol{\psi},\boldsymbol{y})}
\propto \frac{p(\boldsymbol{\theta},\boldsymbol{\psi}|\boldsymbol{y})}{p(\boldsymbol{\theta}_{-i}|theta_i,\boldsymbol{\psi},\boldsymbol{y})} \\
&\approx \frac{p(\boldsymbol{\theta},\boldsymbol{\psi}|\boldsymbol{y})}{\tilde{p}(\boldsymbol{\theta}_{-i}|\theta_i,\boldsymbol{\psi},\boldsymbol{y})}\bigg|_{\boldsymbol{\theta}_{-i}=\boldsymbol{\theta}_{-i}^*(\theta_i,\boldsymbol{\psi})} =: \tilde{p}(\theta_i|\boldsymbol{\psi},\boldsymbol{y}) 
\end{align*}

where $\tilde{p}(\theta_i|\boldsymbol{\psi},\boldsymbol{y})$ is the Gaussian Laplace approximation of $p(\theta_i|\boldsymbol{\psi},\boldsymbol{y})$ around its mode $\boldsymbol{\theta}_{-i}^*(\theta_i,\boldsymbol{\psi})$.

In \texttt{R-INLA} other approximation strategies are implemented and can be chosen to speed up the computation. 

In particular, $p(\theta_i|\boldsymbol{\psi},\boldsymbol{y})$ can be directly derived from the Normal approximation $\tilde{p}(\boldsymbol{\theta}|\boldsymbol{\psi},\boldsymbol{y})$ already computed in the first step (Gaussian strategy). This can produce inaccurate approximations, however when the conditional $p(\theta|\boldsymbol{\psi},\boldsymbol{y})$ is Gaussian it is an exact approximation and there is no need to apply the Laplace method, as in our case. 

For point-referenced data (i.e. data observed at point locations typically referenced by coordinates), the latent continuous spatial process is a Gaussian Field (GF) with dense spatial covariance matrix that leads to computational issues. The SPDE approach proposed by \citet{Lindgren2011} is an alternative which consists in representing a continuous spatial process (the GF) as a discretely indexed spatial random process (i.e. a GMRF).

The continuous GF with Mat\'ern covariance structure $z(\boldsymbol{s})$ is the exact and stationary solution of the following stochastic partial differential equation:
\begin{equation} \label{eq:spde}
(\kappa^2-\Delta)^{\alpha/2}(\tau z(\boldsymbol{s}))=\mathcal{W}(\boldsymbol{s})
\end{equation}
where $\Delta$ is the Laplacian, $\alpha$ is a smoothness parameter, $\kappa$ is a scale parameter, $\tau$ controls the variance, $\boldsymbol{s}$ is the generic spatial location and $\mathcal{W}(\boldsymbol{s})$ is a Gaussian spatial white noise process.

For the relationship between the SPDE in Eq.~(\ref{eq:spde}) and the Mat\'ern parameters see Eq.~(\ref{eq:matern}) in Appendix \ref{app:matern}.

For details on how to solve the SPDE that gives a Mat\'ern random field see \citet{Bakka2018a}.

This solution $z(\boldsymbol{s})$ can be approximated through a weighted sum of basis functions $\psi_g$ defined at the $G$ vertices (nodes) of a triangulation (mesh) of the domain with zero-mean Gaussian-distributed weights $\tilde{z}_g$ \citep{Lindgren2011}: 
\begin{equation*}
z(\boldsymbol{s})=\sum_{g=1}^G \psi_g(\boldsymbol{s})\tilde{z}_g
\end{equation*}
Therefore, at a discrete set of locations $\boldsymbol{s_0}=(\boldsymbol{s}_1, ..., \boldsymbol{s}_G)$ , i.e. the mesh nodes, the GP $\boldsymbol{z}$ follows a multivariate Normal distribution with zero mean and spatially structured correlation matrix.

\begin{figure}
	\centerline{\includegraphics[trim={2cm 2cm 2cm 1.5cm},clip,height=.75\textwidth, angle=-90]{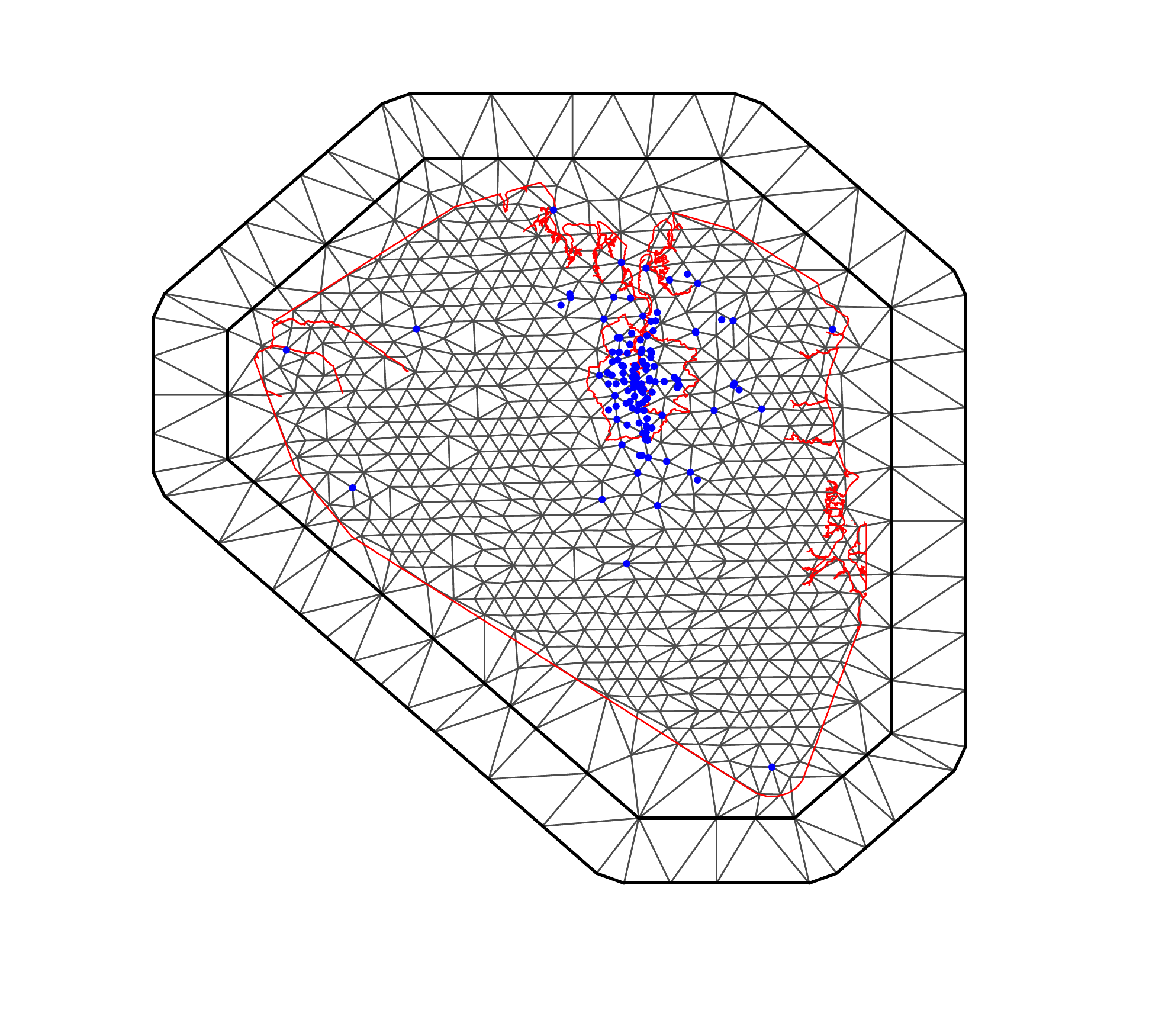}}
	\caption{Domain triangulation (mesh) with 686 vertices (nodes). The red lines are the boundaries of Greater London and the south-eastern coast of UK representing the study area. The blue dots are the locations of the monitoring stations.}
	\label{f:mesh}
\end{figure}

Figure \ref{f:mesh} shows the mesh constructed on our data. 

At the data locations we write $z_i = A z(s_0)$ where $A$ is the projection matrix from the mesh nodes to the data locations. This simplifies the notation as the distribution assumed at the data locations has a complicated form.

\section{Mat\'ern covariance function and INLA-SPDE} \label{app:matern}

Let $r=d(i,j)=||s_i-s_j||$ for two-dimensional domains. Then the Mat\'ern class of stationary and isotropic covariance functions is defined as
\begin{equation} \label{eq:matern}
COV(r) = \dfrac{\sigma^2}{\Gamma(\lambda)2^{\lambda-1}}(\kappa r)^{\lambda}K_{\lambda}(\kappa r)  
\end{equation}

where $\sigma^2 = \dfrac{\Gamma(\lambda)}{\Gamma(\alpha)(4\pi)^{d/2}\kappa^{2\lambda}\tau^2}$ is the marginal variance, $K_{\lambda}$ is the modified Bessel function of second kind and order $\lambda>0$, $\lambda=\alpha-d/2$ is a smoothness parameter, where $d$ is the dimention of the domain (2 in case of spatial processes). We refer the reader to \citet{Lindgren2015} for a discussion on the choice of $\alpha$ implemented in \texttt{R-INLA}. Finally, $\kappa$ is a scaling parameter related to the empirically derived range $\rho=\sqrt{8\lambda}/\kappa$.

We adopt the default $\alpha=2$, which for two-dimensional domains means smoothness parameter $\lambda=1$.

We need a parameterization to represent the Mat\'ern correlation structure in INLA. The easiest way would be to assign a joint normal prior distribution to $\psi_1=log(\tau)$ and $\psi_2=log(\kappa)$, in fact from the formula of the marginal variance $\sigma^2$ we can derive 
\begin{equation} \label{eq:logtau}
log(\tau) = \dfrac{1}{2}log\left(\dfrac{\Gamma(\lambda)}{\Gamma(\alpha)(4\pi)^{d/2}}\right) - log(\sigma) -\lambda log(\kappa)
\end{equation}
and from the formula of the empirical range $\rho=\sqrt{8\lambda}/\kappa$ we obtain 
\begin{equation} \label{eq:logkappa}
log(\kappa) = \dfrac{1}{2}log(8\lambda) - log(\rho)
\end{equation}

A more naturally interpretable parameterization is in terms of standard deviation $\sigma$ and range $\rho$, such as 
\begin{equation} \label{eq:logsigma}
log(\sigma) = log(\sigma_0) + \psi_1
\end{equation}
\begin{equation} \label{eq:logrho}
log(\rho) = log(\rho_0) + \psi_2
\end{equation}

Substituting (\ref{eq:logsigma}) in (\ref{eq:logtau}) and (\ref{eq:logrho}) in (\ref{eq:logkappa}) we get respectively:
\begin{equation} 
log(\tau) = log(\tau_0) -\psi_1 -\lambda log(\kappa)
\end{equation} 
\begin{equation} 
log(\kappa) =log(\kappa_0) - \psi_2
\end{equation}
from which
\begin{equation} \label{eq:logtau1}
log(\tau) = log(\tau_0) -\psi_1 -\lambda \psi_2
\end{equation}
\begin{equation} \label{eq:logkappa1}
log(\kappa) =log(\kappa_0) - \psi_2
\end{equation}
so we can express $\tau$ and $\kappa$ in terms of $\psi_1$ and $\psi_2$ \citep{Lindgren2015}.

More recently, \citet{Fuglstad2017} suggested an extension of the penalized complexity prior (PC prior) proposed by \citet{simpson2017}. The PC prior penalizes the Kullback-Liebler divergence between the model $P$ and the base model $P_0$, i.e. the information lost when approximating $P$ with $P_0$, and it is suggested as a way to reduce overfitting. Here $P$ represents a model component, such as a GRF. 

\citet{Fuglstad2017} developed a joint version of the PC prior for the range and marginal variance of Mat\'ern GRFs with fixed smoothness and $d<4$. The joint PC prior is derived from the alternative parameterization of the Mat\'ern covariance function ($\tau,\kappa$) and then transformed back on the range and variance scale ($\rho,\sigma$). The advantage of PC priors compared to the prior described above is that it can be expressed in a user friendly form stating the upper or lower tail probability for the generic parameter of interest $\phi$: $P(\phi > U) = q$ or $P(\phi < L) = q$. This intuitive interpretation makes it easy to specify a vague, weak or informative prior by tuning the parameter. 

In the case of the joint PC prior for GRFs, we must choose an upper limit for the standard deviation and a lower limit for the range: $P(\sigma > \sigma_0) = q_1$ and $P(\rho < \rho_0) = q_2$.

\section{Implementation of the joint model through the INLA-SPDE approach} \label{a:joint_inla}

Because we have misaligned data, we describe the joint model with three likelihoods and three linear predictors (section \ref{s:model}, equations \ref{eq:linpred1}, \ref{eq:linpred2}, \ref{eq:linpred3}).

To implement this in \texttt{R-INLA}, we need to create a complex data structure: the matrix of observations $M$ is defined as a block matrix with the number of columns corresponding to the number of likelihoods and each block row corresponding to the data used to estimate one of the linear predictors \citep{Martins2013}. 

$$ \boldsymbol{M}=
\begin{pmatrix}
\begin{pmatrix}y_1(1,1) \\\vdots \\ y_1(s_1,t_1) \end{pmatrix} & \textbf{NA} & \textbf{NA}  \\
\textbf{NA} & \begin{pmatrix} y_2(1,1) \\\vdots \\ y_2(s_2,t_2)\end{pmatrix} & \textbf{NA}  \\
\textbf{NA} & \textbf{NA} & \begin{pmatrix}   y_3(1,1)   \\ \vdots \\   y_3(s_3,t_2)  \end{pmatrix}  \\
\end{pmatrix}
$$

The dimension of M is then $(s_1t_1 + s_2t_2 + s_3t_2) \times 3$, with $s_1=44117$ PCM grid cells, $s_2=495$ AQUM grid cells, $s_3=124$ monitoring stations, $t_1=5$ years, $t_2=1826$ days.

Through the \texttt{R-INLA copy} function, the $\boldsymbol{z_1}$ and $\boldsymbol{z_2}$ random effects are included in the linear predictor, so each of them shares the hyperparameters across the linear predictors in eq. \ref{eq:linpred1}-\ref{eq:linpred3}, but at the same time has a scaling parameter as well for calibration purposes \citep{Gomez-Rubio2019a,Rue2016}.

Since the SPDE provides the approximation of the entire spatial process at the mesh nodes, there is no actual alignment procedure involved here.  
For the blocks with spatial structure (all three in our case), we just need a link between the mesh nodes and the locations at which the value is known; this link is provided by a projector matrix defined by built-in functions. Because these locations are different, we need one projector matrix for the PCM data $\boldsymbol{y}_1$, one for the AQUM data $\boldsymbol{y}_2$ and one for the ground observations $\boldsymbol{y}_3$.

For each block, a \texttt{R-INLA} stack object is created to link the data and/or the projector matrix to the model effects included in the linear predictor: for the $\boldsymbol{y}_1$ stack we will have the intercept and only one random effect represented by the spatial index, for $\boldsymbol{y}_2$ we have the intercept, a temporal and a spatial index, and for $\boldsymbol{y}_3$ we will need the intercept, the site type covariate, the spatial index for $\boldsymbol{z}_1$, the temporal index for $\boldsymbol{z}_2$ and the time-site type interaction for $\boldsymbol{z}_3$.

All the stack objects are then put together and passed on to the \texttt{inla} call as data.

With this data structure it is easy to include a validation set: assuming we select $m$ sites for validation and the remaining $n=s_3-m$ for estimation, we just need to set the observations corresponding to the validation locations to $NA$, so that \texttt{R-INLA} will assume them as unknown and will predict their values. In this case the $\boldsymbol{M}$ matrix will be: 
$$ \boldsymbol{M}=
\begin{pmatrix}
\begin{pmatrix}y_1(1,1) \\\vdots \\ y_1(s_1,t_1) \end{pmatrix} & \textbf{NA} & \textbf{NA}  \\
\textbf{NA} & \begin{pmatrix} y_2(1,1) \\\vdots \\ y_2(s_2,t_2)\end{pmatrix} & \textbf{NA}  \\
\textbf{NA} & \textbf{NA} & \begin{pmatrix}   y_3(1,1)   \\ \vdots \\   y_3(n,t_2) \\   NA(n+1,1) \\ \vdots \\  NA(n+m,t_2) \end{pmatrix}  \\
\end{pmatrix}
$$

We also need the projector matrix associated with the validation set and the corresponding stack object. 
The prediction is computed at each new location (either a validation site or on a regular grid) through the procedure described in Section~\ref{s:predictions}.

\section{Retrieving site type at unknown locations} \label{a:sitetype}

To determine the site type at unknown locations, we tested 12 different approaches on the known monitoring sites using four different land cover sources of information to retrieve the rural and urban classification, and three different rules to determine the road-kerb side classification based on the distance from a major or minor road (Greater London Ordnance Survey minor and major roads ESRI shapefile - \citealt{osroads}). 

The first source for land cover is the Corine land cover for the year 2012 for the UK, Jersey and Guernsey shapefile from the Centre for Ecology and Hydrology \citep{ceh}, the second is the MODIS land cover type 1 classification raster at 500m resolution for year 2005 from \citet{modis}, and the last two are the Global Urban Footprint (GUF) rasters at 1km and 12m resolution respectively \citep{dlr, Esch2017}. In all cases, the non-urban land cover types are aggregated to rural, and we assume the land cover did not change significantly over the study period.

We applied the following three rules to all the land cover data: 

(R1) The site type is defined as road-kerb side if the location is within 4 m from any road, otherwise urban or rural according to the land cover shapefile. This rule combined with the MODIS land cover is the one applied by \citet{Mukhopadhyay2017}.

(R2) The site type is defined as road-kerb side if the location is within 10 m from any road, otherwise urban or rural according to the landcover shapefile. 

(R3) The site type is defined as road-kerb side if the location is within 50 m from a major road or within 10 m from a minor road, otherwise urban or rural according to the landcover shapefile. This rule accounts for the different width of the roads, assuming the road midline corresponds to the center of the street.

\begin{table}
	\caption{Accuracy of methods to retrieve site type based on the 126 known monitors.}
	\label{t:sitetype}
	\centering
	\begin{tabular*}{.5\textwidth}{@{}l@{\extracolsep{\fill}}c@{\extracolsep{\fill}}c@{\extracolsep{\fill}}c@{}}
		\hline 							
		Land use& R1	& R2	& R3 	 \\
		\hline
		Corine  	& 43.2\%	& 63.2\%  	& 64.8\%	\\
		MODIS  	& 39.2\%	& 60.8\%  & 64.0\% 	\\
		GUF 1km  	& 37.6\%	& 58.4\%  	& 62.4\%	\\
		GUF 12m  	& 39.2\%	& 60.8\%  	& 63.2\%	\\
		\hline
	\end{tabular*}
\end{table}

The percentage of correctly classified monitors for each method is reported in Table~\ref{t:sitetype}.

The Corine shapefile seems to provide more accurate information compared to the MODIS raster, and combined with the 10/50m rule it gives the highest percentage of correct classification (64.8\%). This method was therefore applied to the unknown locations.

\section{Code} \label{a:code}

The code is available on the GitHub repository \url{https://github.com/cf416/joint_model_no2}.

The data workspace can be requested directly to the corresponding author.

\end{document}